\documentclass[prc,floatfix,showpacs,twocolumn,nofootinbib]{revtex4-1}
\usepackage{xspace}
\usepackage{slashed}
\usepackage{array,multirow}

\usepackage{graphicx}
\usepackage{graphics}
\usepackage{epsfig}
\usepackage{amsfonts}
\usepackage{amsmath}
\usepackage{amssymb}
\usepackage{dsfont}
\usepackage{color}
\usepackage{xcolor}
\usepackage{pstricks}
\usepackage{pifont}
\usepackage{pgfplots}
\usepackage{tikz} 
\usepackage{pgfplotstable}

\usepackage{bm}  %

\newcommand{\beq}{\begin{equation}}
\newcommand{\eeq}{\end{equation}}
\newcommand{\bea}{\begin{eqnarray}}
\newcommand{\eea}{\end{eqnarray}}
\newcommand{\ben}{\begin{eqnarray*}}
\newcommand{\een}{\end{eqnarray*}}

\newcommand{\boldtau}{\mbox{\boldmath $\tau$}}
\newcommand{\boldsigma}{\mbox{\boldmath $\sigma$}}

\renewcommand{\vec}[1]{{\mathbf #1}} 

\newcommand{\bma}{\begin{pmatrix}}
\newcommand{\ema}{\end{pmatrix}}

\def\lixo#1{}

\def\slashchar#1{\setbox0=\hbox{$#1$}           
  \dimen0=\wd0                                    
  \setbox1=\hbox{/} \dimen1=\wd1                  
  \ifdim\dimen0>\dimen1                           
    \rlap{\hbox to \dimen0{\hfil/\hfil}}            
    #1                                             
  \else                                          
    \rlap{\hbox to \dimen1{\hfil$#1$\hfil}}        
    /                                           
 \fi}                                           %

\newcommand{\al}{\alpha}
\newcommand{\bt}{\beta}

\newcommand{\sq}{^{2}}

\newcommand{\dslash}[1]{#1 \llap{/\kern-0.5pt}}
\newcommand{\Dslash}[1]{#1 \llap{/\kern+1.5pt}}
\newcommand{\DDslash}[1]{#1 \llap{/\kern+2.3pt}}
\newcommand{\dslashh}[1]{#1 \llap{/\kern+1pt}}

\newcommand{\Ex}[1]{\cdot 10^{#1}}
\newcommand{\nn}{\nonumber}

\newcommand{\NLDBD}{$0 \nu \beta \beta$}

\begin{document}
\begin{flushright}
LA-UR-17-29297
\end{flushright}
\title{Neutrinoless double beta decay matrix elements in light nuclei}
\author{S. Pastore$^a$,  J. Carlson $^a$, V. Cirigliano$^a$, W. Dekens$^{a,b}$, E. Mereghetti$^a$,  and R.B. Wiringa$^c$}
\affiliation{
$^{\rm a}$\mbox{Theoretical Division, Los Alamos National Laboratory, Los Alamos, NM 87545 }
$^{\rm b}$\mbox{New Mexico Consortium, Los Alamos Research Park, Los Alamos, NM 87544, USA}
$^{\rm c}$\mbox{Physics Division, Argonne National Laboratory, Argonne, IL 60439}
}
\date{\today}

\begin{abstract}

We present the first {\it ab initio} calculations of neutrinoless double beta decay
matrix elements in $A=6$--$12$ nuclei using Variational  Monte Carlo   
wave functions obtained from the Argonne $v_{18}$  two-nucleon potential and 
Illinois-7  three-nucleon interaction. 
We study both light Majorana neutrino exchange  and  potentials arising from a large class 
of multi-TeV mechanisms of lepton number violation.  
Our results provide 
benchmarks to be used in testing  many-body 
methods that can be extended to the heavy nuclei of experimental interest.   
In  light nuclei   
we have also studied the impact of two-body short range correlations 
and the use of different forms for the transition operators,  
such as those  corresponding to different orders in chiral effective theory.

\end{abstract}

\maketitle

\section{Introduction}

Searches for  neutrinoless double beta decay ($0\nu\beta\beta$)   
constitute the most sensitive  laboratory probe of lepton number violation (LNV).
In \NLDBD\  two neutrons in a nucleus turn into two protons, with the emission of 
two electrons and no neutrinos, violating $L$ by two units.
The observation of $0\nu\beta\beta$  would demonstrate that neutrinos are 
Majorana fermions~\cite{Schechter:1981bd},  shed light on the mechanism of neutrino mass generation, 
and give insight into leptogenesis scenarios for the generation of the matter-antimatter 
asymmetry in the universe~\cite{Davidson:2008bu}. 

For certain even-even nuclei the  single $\beta$ decay is energetically forbidden. 
In  many such nuclei,  the Standard Model allowed  two-neutrino double beta decay has 
already been observed \cite{Argyriades:2008pr,Argyriades:2009ph,Ackerman:2011gz,Agostini:2012nm,KamLANDZen:2012aa,Alduino:2016vtd} (see Ref. \cite{Saakyan:2013yna} for older references),  
and the search for the  LNV neutrinoless mode is being pursued by many collaborations worldwide. 
The current experimental  limits on the half-lives for the neutrinoless mode are  quite  
impressive~\cite{KamLAND-Zen:2016pfg, Alfonso:2015wka,Albert:2014awa,Agostini:2013mzu,Gando:2012zm,Elliott:2016ble,Andringa:2015tza,Agostini:2017iyd},
at the level  of $T_{1/2} >  5.3\times10^{25}$~y  for $^{76}$Ge~\cite{Agostini:2017iyd}
and  $T_{1/2} >  1.07\times10^{26}$~y  for $^{136}$Xe~\cite{KamLAND-Zen:2016pfg}, 
with next generation ton-scale experiments aiming at two orders of magnitude sensitivity improvements.

The observation of \NLDBD,  while of great significance by itself,  
would not  immediately point to the underlying  mechanism of  lepton number violation. 
In fact, next-generation experiments are sensitive to a variety of mechanisms,  which are most  efficiently  
discussed in an effective theory approach to new physics, in which LNV arises from $\Delta L=2$ operators of odd dimension, 
starting at dimension-five~\cite{Weinberg:1979sa,Babu:2001ex,deGouvea:2007qla,Lehman:2014jma}.
As discussed for example in Ref.~\cite{Cirigliano:2017djv}, 
if the scale of  lepton number violation, $\Lambda_{\rm LNV}$ is in the range 1-100 TeV,  
short-distance  effects  encoded in local operators of  dimension seven and nine 
provide  contributions to \NLDBD\ within reach of next generation experiments.  
On the other hand, whenever  $\Lambda_{\rm LNV}$ is much higher than the TeV scale,  
the only low-energy manifestation of this new physics is a Majorana mass 
for light neutrinos, encoded in a  single gauge-invariant dimension-five operator~\cite{Weinberg:1979sa}, 
which induces \NLDBD \ through light Majorana-neutrino exchange~\cite{Bilenky:2014uka,Bilenky:1987ty}.

To interpret positive or null \NLDBD \ results in the context of various LNV mechanisms  
it is  essential  to have  control over the  relevant hadronic and nuclear matrix elements. 
Current knowledge of these is somewhat  unsatisfactory~\cite{Engel:2016xgb}, as various many-body 
approaches lead to estimates that differ by a factor of two to three for nuclei of experimental interest.   
This is true both for the  light Majorana-neutrino exchange  mechanism, which  has received much attention in the literature, 
and for  short-distance sources of LNV encoded in dimension-seven and -nine operators (see~\cite{Cirigliano:2017djv} and 
references therein). 

In this paper we present the first {\it ab initio} calculations of  \NLDBD\ 
nuclear matrix elements in light nuclei  ($A=6$--$12$),  using Variational Monte Carlo (VMC) 
wave functions obtained from the Argonne $v_{18}$ (AV18)~\cite{AV18}  two-body potential and 
Illinois-7  (IL7)~\cite{IL7} three-nucleon interaction. 
We use the measured value of the axial coupling constant
$g_A=1.2723(23)$~\cite{Olive:2016xmw}---also utilized in recent {\it ab initio} quantum Monte 
Carlo calculations of 
single beta decays in $A=6$--$10$ nuclei~\cite{Pastore17} that explain the data at 
the $\leq 2\%$ ($\sim 10\%$) level in $A=6$--$7$ ($A=10$) decays---and compare
with results for $A=48$--$136$ nuclei~\cite{Javier,Hyvarinen:2015bda}
also based on the measured value of $g_A$.
We study the matrix elements of light Majorana-neutrino exchange 
as well as those arising from a large class of multi-TeV mechanisms of LNV.  
While the transitions studied here are not directly relevant from 
an experimental point of view,  this study has several merits:  
(i)  Because  the  {\it ab initio} framework used here accurately
explains, qualitatively and quantitatively, the observed properties  
of light nuclei~\cite{Carlson15,Carlson98,review2014}, 
our results provide  an important benchmark  to test other many-body 
methods that can be extended to the heavy nuclei of experimental interest.
(ii) In this framework we can study  in a controlled way the impact 
of various approximations  inherent to some many-body 
methods -- such as neglecting two body correlations.
(iii)  For a given LNV mechanism, we can explore the impact of using  
different forms for the transition operators (``potentials") mediating \NLDBD. 
(iv) In the same vein, we can  study  the relative size of 
matrix elements corresponding to different LNV mechanisms.

The paper is organized as follows.  In Section~\ref{sect:potentials} we present the two-body  transition operators (``potentials") 
that mediate \NLDBD \  from a large class of LNV mechanisms. 
In Section~\ref{sec:vmc} we describe the VMC method and in Section~\ref{sec:results} we discuss our results.
We present our conclusions in Section~\ref{sect:concl} and provide some details on the potentials in 
coordinate space in Appendix~\ref{App1}.

\begin{figure}[bth]
\centering
 \includegraphics[width=2.5in]{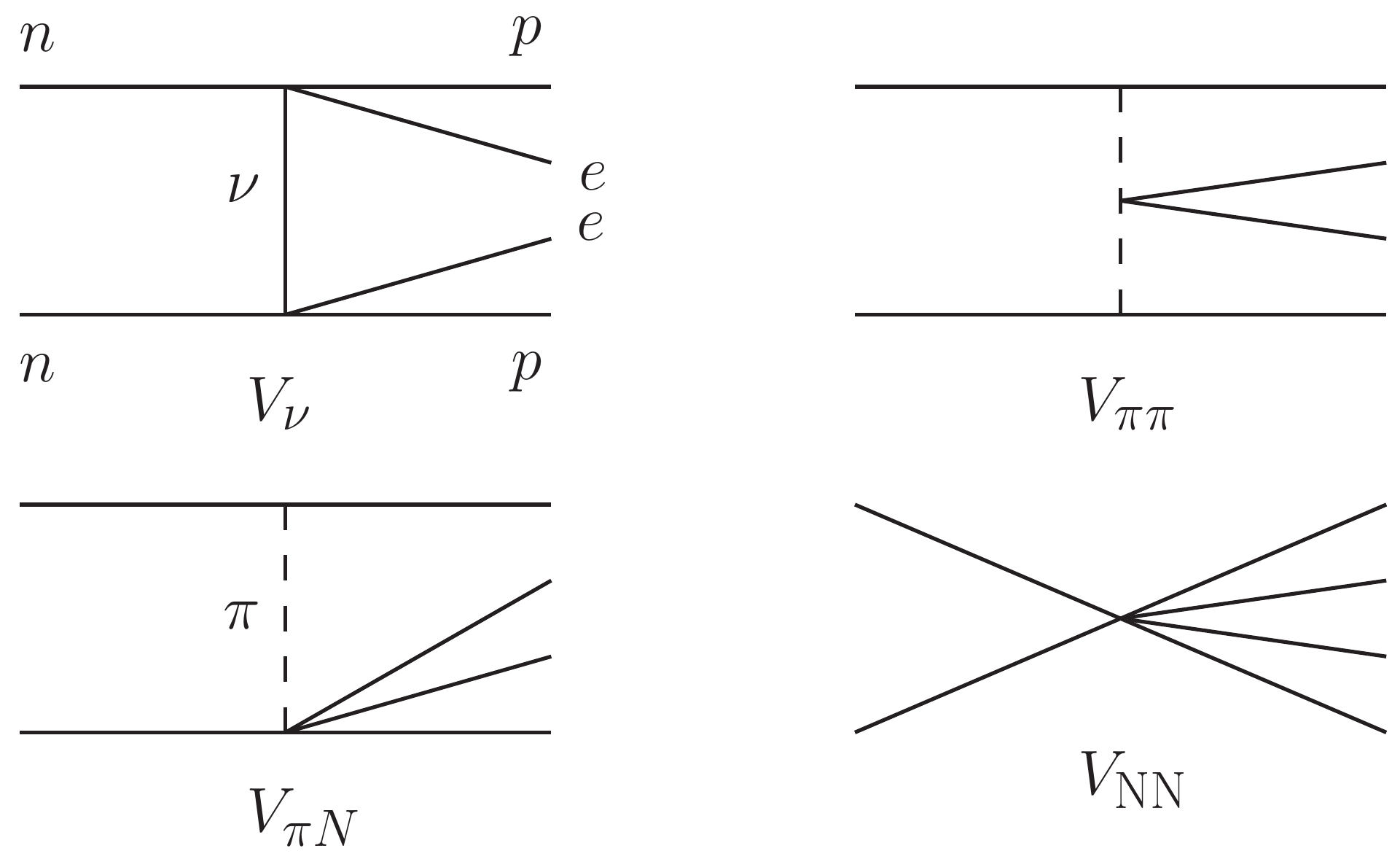} 
 \caption{Diagrams illustrating the $0\nu\beta\beta$ potentials mediated by 
 neutrinos---$V_\nu$ defined in Eq.~(\ref{eq:vnu})---and two-pion-exchange, one-pion-exchange,  
 and short-distance interactions---$V_{\pi\pi}$, $V_{\pi N}$, and $V_{NN}$
 defined in Eqs.~(\ref{eq:dim9pot}). }
 \label{fig:diag}
\end{figure}

\section{Nuclear operators for \NLDBD}
\label{sect:potentials}

\subsection{Matching quark  operators to hadronic operators}

Our starting point is  a  $\Delta L = 2$  effective Lagrangian
${\cal L}_{\Delta L = 2}$    at the hadronic scale $E \sim \Lambda_\chi \sim$~GeV
written in terms of leptons and  quarks. 
This effective Lagrangian   originates from integrating out heavy new physics at the scale $\Lambda_{\rm LNV}$ and matching onto  
$SU(3)_C \times SU(2)_L \times U(1)_{Y}$-invariant operators. 
After integrating out the heavy SM fields at the electroweak scale, one obtains a set of $SU(3)_C \times U(1)_{\rm EM}$-invariant operators that we incorporate into our effective Lagrangian.
In this work, with the purpose of benchmarking nuclear matrix elements,  
we include 
only the dimension-three Majorana neutrino mass operator 
and a subset of dimension-nine  six-fermion operators that  mediate   short-range contributions to \NLDBD:  
\bea
\label{eq:Leff1}
\mathcal L_{\Delta L=2} &=&  -  \frac{1}{2}   m_{\beta \beta}  \ \nu^T_{eL} \, C \nu_{eL}  
+  \mathcal L^{(9)}_{\Delta L = 2}+{\rm h.c.}\,\,,
\\
 \mathcal L^{(9)}_{\Delta L = 2} & = & 
\frac{V_{ud}^2}{v^5}  \times  \bar e_{L} C \bar e_{L}^T
 \bigg\{
C^{(9)}_{1}\,\bar u_L \gamma^\mu d_L\,  \bar u_L \gamma_\mu d_L 
\label{eq:Ldim9}
\\
&+&  C^{(9)}_{2}\,\bar u_L   d_R\,  \bar u_L   d_R   
+C^{(9)}_{3}\,\bar u_L^\al   d_R^\bt \,  \bar u_L^\bt   d_R^\al   
\nonumber \\
&+&C^{(9)}_{4}\, \bar u_L \gamma^\mu d_L\,  \bar u_R \gamma_\mu d_R
+C^{(9)}_{5}\, \bar u^\al_L \gamma^\mu d^\bt_L\,  \bar u^\bt_R \gamma_\mu d^\al_R
\bigg\}~.
\nonumber
\eea
Here $v = ( \sqrt{2} G_F)^{-1/2} = 246$~GeV,  $\al$, $\bt$ are  color indices, 
and for later convenience we have extracted a factor of 
$V_{ud}^2 $ 
from the dimensionless Wilson coefficients $C_i^{(9)}$.
The dimension-three term in Eq.\ \eqref{eq:Leff1} originates from the only $SU(2)_L$-invariant operator at dimension-five, while the dimension-nine terms can arise from both dimension-seven and -nine $SU(2)_L$-invariant operators.

In principle, the most general $\Delta L=2$ low-energy effective Lagrangian would include additional dimension-six and -seven charged-current operators,
which give rise to long-range contributions to \NLDBD, not proportional to $m_{\beta\beta}$. However, as was shown in Ref.~\cite{Cirigliano:2017djv},  the nuclear matrix elements (NMEs)
needed in this case are related to NMEs that appear in  light and heavy Majorana-neutrino exchange, and thus do not require independent calculations.
Furthermore, the effective Lagrangian  in \eqref{eq:Ldim9} represents a subset of the most general  
dimension-nine $\Delta L=2$ interactions.
The complete basis of dimension-nine operators includes additional terms that can be obtained by the 
interchange of $L \leftrightarrow R$  on the quark and/or lepton-fields in  \eqref{eq:Ldim9}, as well as operators  
in which the quark and electron structures are Lorentz vectors (e.g.  $\bar{e}_L \gamma_\mu C \bar{e}_R^T$) \cite{Graesser:2016bpz,Prezeau:2003xn}. However, 
as far as $0^+ \rightarrow 0^+$ transitions are concerned, none of these additional operators lead to different hadronic realizations than those induced by the operators in Eq.\ \eqref{eq:Ldim9} \cite{Cirigliano:2017ymo}. As a result, the NMEs studied in the following capture the leading contributions to \NLDBD\ from $SU(2)_L$-invariant operators of dimension-five and -seven, as well as those from dimension-nine operators involving six fermions.

The leading low-energy realization of the effective Lagrangian \eqref{eq:Leff1} in terms of
leptons, pions, and nucleons, reads~\cite{Cirigliano:2017ymo,Prezeau:2003xn}
\begin{widetext}
\begin{eqnarray}
\mathcal L_{\Delta L = 2} &=&
-  \frac{1}{2}   m_{\beta \beta}  \ \nu^T_{eL} \, C \nu_{eL}   
+   \frac{V_{ud}^2}{v^5}  \times  \bar e_{L} C \bar e_{L}^T
\left\{  \frac{5}{6} \, C_1^{(9)} \, g_{27 \times 1}\,  F_\pi^2 \partial_\mu \pi^- \partial^\mu \pi^-  
 \nonumber \right. \\ &  & \left. 
+ \frac{1}{2} F_\pi^2 \left(
 C^{(9)}_{4} g_{8 \times  8} + C^{(9)}_{5} g^{\textrm{mix}}_{8 \times  8}
 -  C^{(9)}_{2} g_{6 \times  6} - C^{(9)}_{3} g^{\textrm{mix}}_{6 \times  \bar 6} 
\right)  \pi^- \pi^-  
 \nonumber \right. \\ &  & \left. 
+ \sqrt{2} g_A F_\pi 
 C_1^{(9)}  \, g^{\pi N}_{27\times 1}\   \bar p S \cdot (\partial \pi^-) n 
+ \frac{1}{2} C_1^{(9)}   g^{NN}_{27\times 1} \    \bar p  n \, \bar p n
\right\}  \ \,\,.
\label{eq:Leff2}
\end{eqnarray}
\end{widetext}
The low-energy constants (LECs) $g_{8 \times 8}$ and $g_{6 \times \bar{6}}$ are of $\mathcal O(\Lambda_\chi^2)$, 
while  $g_{27\times1}$ and $g^{\pi N}_{27\times1}$  are of $\mathcal O(1)$. The coupling constant of 
the $\Delta L=2$ four-nucleon operator, $g^{NN}_{27\times1}$, is $\mathcal O(1)$ in the Weinberg power counting \cite{Weinberg:1990rz,Weinberg:1991um}.
We follow the notation of Ref.~\cite{Cirigliano:2017ymo}, in which  $g_{8 \times 8}$,  $g_{6 \times \bar 6}$,  and $g_{27 \times 1}$ 
(see also Ref.~\cite{Savage:1998yh})  were estimated using $SU(3)$ chiral perturbation theory ($\chi$PT) relations and lattice-QCD calculations of kaon matrix elements. 
At $\mu = 3$~GeV in the $\overline{\rm MS}$ scheme  one has  $g_{27 \times 1} =  0.37 \pm 0.08$,
$g_{8 \times 8} = -(3.1 \pm 1.3)$~GeV$^2$, 
$g^{\textrm{mix}}_{8 \times 8}=  -(13 \pm 4)$~GeV$^2$, 
$g_{6 \times \bar 6} =  (3.2 \pm  0.7)$~GeV$^2$, 
$g^{\textrm{mix}}_{6 \times \bar 6}=  -(1.1 \pm 0.3)$~GeV$^2$.
For the new-physics operators that transform as $8_L \times 8_R$ or $6_L \times \bar 6_R$, within the Weinberg power counting, only the $\pi\pi$  interactions contribute at LO, and we neglect the subleading pion-nucleon and nucleon-nucleon couplings in Eq.\ \eqref{eq:Leff2}. Instead, for the operator transforming as $27_L\times 1_R$, we include all three types of interactions as they contribute to \NLDBD\ at the same order.

\subsection{The isotensor nuclear potentials}

From  the effective Lagrangian \eqref{eq:Leff2}  one obtains the following 
$\Delta L = 2$ effective hamiltonian for \NLDBD\   in terms of electrons and nucleons: 
\begin{equation}
H_{\Delta L = 2} =  2 G_F^2\, V_{ud}^2\,  \bar e_{L} C \bar e_{L}^T \ \   \sum_{a,b}  V  (a,b) \,\,,
\end{equation}
with the isotensor  potential  given by 
\begin{eqnarray}
V   &=&    m_{\beta \beta}   \, V_\nu   +    \frac{m_\pi^2}{v}  \left( 
c_{\pi \pi}  V_{\pi \pi}   +  c_{\pi N}  V_{\pi N}  + c_{NN}  V_{NN} \ 
\right). \ \ 
\label{LOpot}
\end{eqnarray}
In what follows we will give the two-body potentials in momentum space, 
while providing their coordinate space expressions in  Appendix~\ref{App1}.

\subsubsection{Light Majorana neutrino exchange}
The first term in Eq.~\eqref{LOpot} is generated by light Majorana-neutrino exchange, depicted in the top-left panel of Fig.\ \ref{fig:diag}, and at leading order is given by 
\begin{eqnarray}
\label{eq:vnu}
V_\nu  &=&  \tau^{+}_a \tau^{+}_b  
\,  \frac{1}{\vec{q}^2}  \,  \Bigg\{
g_V^2 \nonumber \\
&-&  g_A^2  \Bigg[
 \boldsigma_{a} \cdot \boldsigma_{b} \left(1 - \frac{2}{3} \frac{\vec q^2}{\vec q^2 + m_\pi^2}  
 + \frac{1}{3} \frac{(\vec q^2)^2}{(\vec q^2 + m_\pi^2)^2} \right) 
 \nonumber 
\\
&- & 
\frac{S_{ab} (\hat{\bf q})}{3}  \left( - \frac{2 \vec q^2}{\vec q^2 + m_\pi^2} +  \frac{(\vec q^2)^2}{(\vec q^2 + m_\pi^2)^2} \right) 
 \Bigg]   \Bigg\}~, 
\end{eqnarray}
where $\hat{\vec q} = \vec q/|\vec q|$, $g_V =1$, $g_A = 1.27$, and the tensor operator  is given by 
$S_{ab}= -\left( 3\,\boldsigma_{a} \cdot  \hat{ \vec q} \, \boldsigma_{b} \cdot  \hat{ \vec q} - \boldsigma_{a}\cdot \boldsigma_{b} \right)$ in momentum space.
Higher-order corrections to the single-nucleon charged-currents can be taken into
account by including momentum-dependent form factors. Here we follow Ref. \cite{Engel:2016xgb} and express $V_\nu$ as
\begin{eqnarray}\label{Vnu}
V_\nu &=& \tau^{+}_a \tau^{+}_b  
\,  \frac{g_A^2}{\vec{q}^2}   \, \Bigg\{  \frac{g_V^2}{g_A^2} v_F^\nu(\vec q^2) \nonumber \\
&-&  \boldsigma_a \cdot \boldsigma_b  \, v_{GT}^\nu(\vec q^2)   
-   S_{ab}\,  v^\nu_T(\vec q^2) \Bigg\}.
\end{eqnarray}
The Fermi (F), Gamow-Teller (GT) and tensor (T) functions can be expressed in terms of the nucleon isovector vector, 
axial, induced pseudoscalar and tensor form factors as 
\begin{eqnarray}\label{eq:hK(q)}
&& v_F^\nu(\vec q^2) =   g_V^2(\vec q^2)/g_V\sq\,,\nn\\ 
&& v_{GT}^\nu(\vec q^2) = v^{AA}_{GT}(\vec q^2) + v^{AP}_{GT}(\vec q^2) + v^{PP}_{GT}(\vec q^2) + v^{MM}_{GT}(\vec q^2)\,,\nn\\
&& v_{T}^\nu(\vec q^2)  =  v^{AP}_{T}(\vec q^2) + v^{PP}_{T}(\vec q^2) + v^{MM}_{T}(\vec q^2)\,,
\end{eqnarray}
where for the GT and T terms we have 
\begin{eqnarray}\label{smff}
&& v^{AA}_{GT,T}(\vec q^2) =  \frac{g_A^2(\vec q^2)}{g_A\sq}\,,\nonumber \\
&& v_{GT}^{AP}(\vec q^2) = \frac{g_P(\vec q^2)}{g_A\sq}\, g_A(\vec q^2) \frac{\vec q^2}{3 m_N}\,, \nonumber \\
&& v_{GT}^{PP}(\vec q^2) = \frac{ g^2_P(\vec q^2)}{g_A\sq} \frac{\vec q^4}{12 m_N^2}\,,  \nonumber \\
&& v^{MM}_{GT}(\vec q^2) =  g_M^2(\vec q^2) \frac{\vec q^2}{6g_A\sq m_N^2}\,,
\end{eqnarray}
and $v^{AP}_T(\vec q^2) = -v^{AP}_{GT}(\vec q^2)$, 
$v^{PP}_T(\vec q^2) = - v^{PP}_{GT}(\vec q^2)$, and  $v^{MM}_{T}(\vec q^2) = v^{MM}_{GT}(\vec q^2)/2$.

As commonly done in the \NLDBD\ literature, we use a dipole parameterization for 
the vector and axial form factors, and write 
\begin{eqnarray}\label{eq:FF}
&& \!\!g_V(\vec q^2)\!\! = \!\!g_V \left(1 + \frac{\vec q^2}{\Lambda_V^2}\right)^{-2}, \,\, g_M(\vec q^2)\!\! =\!\! (1 + \kappa_1) g_V(\vec q^2) \,\,,\nonumber \\
&& \!\!g_A(\vec q^2)\!\! = \!\!g_A \left(1 + \frac{\vec q^2}{\Lambda_A^2}\right)^{-2}, \,\, g_P(\vec q^2)\!\! =\!\! -\frac{2 m_N g_A(\vec q^2)}{\vec q^2 + m_\pi^2},
\end{eqnarray}
where the vector and axial masses are $\Lambda_V = 850$ MeV and $\Lambda_A = 1040$ MeV, and the anomalous nucleon isovector magnetic moment $\kappa_1 = 3.7$. 
In the limit $\Lambda_{A,V} \rightarrow \infty$, Eq.\ \eqref{eq:FF} reduces to the leading order (LO) $\chi$PT expression.
In what follows, we define the neutrino potentials in momentum space as 
\bea
V_{\al,\bt}(\vec q^2) = \frac{1}{\vec q^2} v^{\bt}_{\al}(\vec q^2)\,\,,
\eea
with $\al \in \{ F, GT, T\}$ and $\bt \in \{\nu, AA, AP, PP, MM\}$,
and the functions $v_\al^\bt$ given in Eqs.\ \eqref{eq:hK(q)} and \eqref{smff}. 
The potential $V_{T,\, AA}$ does not appear in the case of light Majorana-neutrino exchange, but it is relevant in the presence of 
right-handed charged-currents \cite{Doi:1985dx,Muto:1989cd,Cirigliano:2017djv} 

Non-factorizable contributions to $V_\nu$ 
arise at the same order as form-factor corrections, as recently shown in Ref.~\cite{Cirigliano:2017tvr}. 
We explore the impact of these in Section.~\ref{Sect:Vnu2}.

\subsubsection{LNV from short-distance}
The dimension-nine operators with couplings  $C_i^{(9)}$ induce the pion-range and short-range 
potentials $V_{\pi\pi}$, $V_{\pi N}$ and $V_{NN}$  in Eq.~\eqref{LOpot} through the diagrams shown in Fig.\ \ref{fig:diag}: 
\begin{eqnarray}\label{eq:dim9pot}
&& V_{\pi \pi} = \tau^+_a \tau^+_b  
\left(   \boldsigma_a \cdot \boldsigma_b  - S_{ab}  \right)  \frac{\vec q^2}{3 (\vec q^2 + m_\pi^2)^2}  \ ,\nonumber
\\
&& V_{\pi N} = - \tau^+_a \tau^+_b  
\left(\!   \boldsigma_a \cdot \boldsigma_b     + S_{ab} \frac{\vec q^2}{m_\pi^2}  \! \right) \frac{1}{ 3 (\vec q^2 + m_\pi^2)} \ ,
 \nonumber \\
&& V_{NN } = \tau^+_a \tau^+_b \frac{1}{m_\pi^2}\ .
\end{eqnarray}
As for the light Majorana-neutrino exchange potential $V_\nu$, 
we split the $V_{\pi \pi}$ and $V_{\pi N}$ in  Gamow-Teller 
and tensor components (see Appendix A).
The dimensionless  effective couplings are given by:
\bea
c_{\pi \pi} &=& - \frac{g_A^2}{2 m_\pi^2} \Big( C^{(9)}_{4} g_{8 \times  8} + C^{(9)}_{5} g^{\textrm{mix}}_{8 \times  8}
 -  C^{(9)}_{2} g_{6 \times  6} \nonumber \\
 &-& C^{(9)}_{3} g^{\textrm{mix}}_{6 \times  \bar 6}  + \frac{5}{3} \, C_1^{(9)} \, g_{27 \times 1} m_\pi^2 \Big) \ ,
\\
c_{\pi N} &=& - g_A^2   \, C^{(9)}_1  \, \left(  g^{\pi N}_{27 \times 1}  - \frac{5}{6} \, g_{27 \times 1} \right) \ ,
\\
c_{NN} &=& -  C^{(9)}_1  \left( \, g^{NN}_{27 \times 1} - g_A^2   \,  \left(  g^{\pi N}_{27 \times 1}  
- \frac{5}{6} \, g_{27 \times 1} \right) \right)\ .
\eea
At leading order in chiral EFT,  the potentials in Eq. \eqref{eq:dim9pot} do not include momentum dependent form factors. 
Note that, after absorbing the short-distance pieces of the $c_{\pi N}$ and $c_{\pi\pi}$ contributions into $V_{NN}$, 
we have $V_{GT,\pi\pi}= -V_{GT,PP}$ and $V_{GT,\pi N} = -V_{GT,AP}/2$ (see Appendix \ref{App1}).
In our analysis, we will study the sensitivity to the large momentum region by multiplying $V_{\pi\pi}$, $V_{\pi N}$ and $V_{NN}$ 
by  a dipole form factor, for which we take  $g_A^2(\vec q^2)/g_A^2$.

\subsection{Matrix elements}

To make contact with the standard $0\nu\beta\beta$ literature, it is convenient to define 
the dimensionless matrix elements between the initial and final nuclear states, $|\Psi_i\rangle$
and $|\Psi_f\rangle$, as
\begin{equation}\label{MEdef}
M^{\alpha,\beta} = \langle \Psi_f |  O^{\alpha,\beta}  | \Psi_i \rangle \  , 
\end{equation}
where the two-body F, GT, and T operators are given by
\begin{eqnarray}
\label{eq:opF}
 && O^{F,\beta} 
   =\left(4 \pi  R_A\right)\,  \sum_{a,b} V_{F, \beta}(r_{ab})\, \tau^+_a \tau^+_b \ , \\
 && O^{GT,\beta} 
  =\left(4 \pi  R_A\right)\,  \sum_{a,b} V_{GT,\beta}(r_{ab})\,    \boldsigma_a \cdot \boldsigma_b \, \tau^+_a \tau^+_b \ , \\
 && O^{T,\beta}
 =\left(4 \pi  R_A\right)\,  \sum_{a,b} V_{T, \beta}(r_{ab})\,S_{ab} \, \tau^+_a \tau^+_b \ ,
\label{eq:opT}
 \end{eqnarray}
where $R_A = 1.2\, A^{1/3}$ fm is  the nuclear radius
and now $\beta \in \{ \nu, AA, AP, PP, MM, \pi \pi, \pi N, NN  \}$.
Note that the operators defined above involve an unconstrained sum over $a\neq b$. 
The potentials in momentum and coordinate space are related by 
\begin{equation}
V_{\al,\bt}  (r_{ab})= \int \frac{d^3 q}{(2\pi)^3} e^{i \vec q \cdot \vec r_{ab}} \, V_{\al,\bt}(\vec q).
\end{equation}
For completeness, we report explicit expressions for
the potentials in coordinate space in Appendix \ref{App1}.

\section{Variational Monte Carlo Method}
\label{sec:vmc}

The evaluation of the matrix elements defined in 
Eq.~(\ref{MEdef}) is carried out using 
Variational Monte Carlo (VMC) computational
algorithms~\cite{Carlson15}.  
The VMC wave function $\Psi(J^\pi;T,T_z)$---where $J^\pi$
and $T$ are the spin-parity and isospin of the state---is 
constructed from products of two- and three-body correlation 
operators acting on an antisymmetric single-particle state of 
the appropriate quantum numbers.
The correlation operators are designed to reflect the influence of the
two- and three-body nuclear interactions at short distances, while 
appropriate boundary conditions are imposed at long range~\cite{Wiringa91,Pudliner97}.

The $\Psi(J^\pi;T,T_z)$ has embedded variational parameters
that are adjusted to minimize the expectation value
\begin{equation}
 E_V = \frac{\langle \Psi | H | \Psi \rangle}
            {\langle \Psi   |   \Psi \rangle} \geq E_0 \ ,
\label{eq:expect}
\end{equation}
which is evaluated by Metropolis Monte Carlo integration~\cite{Metropolis53}.
In the equation above, $E_0$ is the exact lowest eigenvalue of the nuclear 
Hamiltonian $H$ for the specified quantum numbers.
The many-body Hamiltonian is given by
\begin{equation}
 \label{eq:nucH} 
 H = \sum_{i} K_i + {\sum_{i<j}} v_{ij} + \sum_{i<j<k} V_{ijk} \ , 
\end{equation}
where $K_i$ is the non-relativistic kinetic energy of nucleon $i$ 
and $v_{ij}$ and $V_{ijk}$ are, respectively, 
the Argonne $v_{18}$ (AV18)~\cite{AV18} 
two-body potential and the Illinois-7
(IL7)~\cite{IL7} three-nucleon interaction. 
The AV18+IL7 model reproduces the experimental
binding energies, charge radii, electroweak transitions
and responses of $A=3$--$12$ systems in numerically exact 
calculations based on Green's function Monte Carlo (GFMC) 
methods~\cite{Carlson98,review2014,Carlson15,Pastore17}. 

A good variational wave function, that serves as the starting
point of GFMC calculations, can be constructed with
\begin{equation}
   |\Psi_V\rangle =
      {\cal S} \prod_{i<j}^A
      \left[1 + U_{ij} + \sum_{k\neq i,j}^{A}\tilde{U}_{ijk} \right]
      |\Psi_J\rangle.
\label{eq:psit}
\end{equation}
The Jastrow wave function $\Psi_J$ is fully antisymmetric, translationally
invariant, and has the
$(J^\pi;T,T_z)$ quantum numbers of the state of interest, while $U_{ij}$
and $\tilde{U}_{ijk}$ are the two- and three-body correlation
operators, and $\cal S$ is a symmetrization operator.
The two-body correlation operators~\cite{Carlson15} can 
be schematically written as 
\begin{equation}
 U_{ij} = \sum_p f^p(r_{ij}) \, O^p_{ij} \ , 
\end{equation}
where 
\begin{equation}
O^p_{ij}= \boldtau_i\cdot \boldtau_j\,,\, \boldsigma_i\cdot\boldsigma_j\,,\,
(\boldtau_i\cdot\boldtau_j)(\boldsigma_i\cdot\boldsigma_j)\,,\,S_{ij}\,
,\,S_{ij}\boldtau_i\cdot \boldtau_j\ ,
\end{equation}
are the main static operators that appear in the two-nucleon potential
and the $f^p$ are functions of the interparticle distance $r_{ij}$
generated by the solution of a set of coupled differential equations 
containing the bare two-nucleon potential with asymptotically-confined 
boundary conditions~\cite{Carlson15}.
In order to study how correlations in the nuclear wave functions
impact on the calculated matrix elements, we perform a calculation
in which we turn off the ``one-pion-exchange-like'' correlation 
operators, {\it i.e.}, $(\boldtau_i\cdot\boldtau_j)(\boldsigma_i\cdot\boldsigma_j)\,$
and $S_{ij}\boldtau_i\cdot \boldtau_j$. The effects such an 
artificial change will be discussed in Sec.~\ref{sec:results}.

In principle, the variational wave function can be further improved
via an imaginary time propagation of the Schr\"odinger equation.
This procedure has the effect of eliminating spurious contributions 
coming from excited states and it is implemented by the GFMC 
algorithm~\cite{Carlson15}. However, Quantum Monte Carlo studies of 
electroweak matrix elements in low-lying nuclear states of 
$A\leq10$ nuclei indicate that the GFMC propagation improves
the VMC results by $\lesssim 3\%$~\cite{Pastore12,Pastore17},
an accuracy that goes beyond the scope of the present investigation.

The results presented below for $A \leq 10$ nuclei use the VMC wave functions
that serve as starting trial functions for the GFMC calculations summarized in 
Ref.~\cite{Carlson15}.
For the $A=12$ nuclei, we use new clusterized variational wave functions that 
provide for alpha- and dineutron-like clusters among the p-shell nucleons.
As for the lighter nuclei, they are fully antisymmetric $A$-body wave 
functions, translationally invariant, and include the same product of two-
and three-body operator correlations induced by the nuclear Hamiltonian.
However, for simplicity, only the highest spatial symmetry states are used, 
i.e., [444] in $^{12}$C and [4422] in $^{12}$Be, as specified in Young
diagram notation~\cite{Wiringa06}.
The construction of $^{12}$C can be thought of as coupling a core $^8$Be 
nucleus in one of its first three states ($0^+$, $2^+$, or $4^+$) with an
additional p-shell alpha-like cluster in respectively a $^1$S$_0$, $^1$D$_2$,
or $^1$G$_4$ state, to give a total $J^\pi = 0$.  
Similarly, for $^{12}$Be, a core $^8$He nucleus in one of its first two
states ($0^+$ or $2^+$) is coupled with a $^1$S$_0$ or $^1$D$_2$ p-shell
alpha-like cluster.
In both cases a small-basis diagonalization is made among these components.
These $A=12$ calculations are computationally demanding because of the
size of the spin-isospin vectors needed to represent the wave function:
4,096 x 132 for $^{12}$C and 4,096 x 275 for $^{12}$Be, where we assume
pure $T=0$ and $T=2$ states, respectively.
This is the first quantum Monte Carlo wave function for $^{12}$Be.

In addition to  presenting results on the matrix elements of Eq.~(\ref{MEdef}),
we study their associated  transition distributions in $r$-space, 
$C^{\alpha,\beta}(r)$, and $q$-space, $\bar{C}^{\alpha,\beta}(q)$
defined as 
 \begin{eqnarray}
\label{eq:densities}
 M^{\al,\beta} &=& \int d{\bf r}\, \rho^{\al,\beta}(r) 
 \equiv \int dr \, C^{\al,\beta}(r) \equiv \int dq \, \bar{C}^{\al,\beta}(q) \ , \nn\\
\end{eqnarray} 
where  $\rho^{\alpha,\beta}(r)$ is the transition density associated
with the transition operator $O^{\alpha,\beta}(r)$.

Finally,  following Ref.~\cite{NV}  we  
represent the delta-functions entering the $V_{GT, MM}$ and $V_{F, NN}$
potentials defined in  Eqs.~(\ref{ourdef.d}) and (\ref{ourdef2})
with 
\begin{equation}
\label{eq:cutoff}
 \delta(m_\pi {\bf r}) =\frac{e^{-(r/R_S)^2}}{m_\pi^3\,R_S^3\, \pi^{3/2}} \ ,
\end{equation}
where $R_S$  is a short range cutoff.  
We tested the sensitivity of
the calculated matrix elements with respect
to variations of $R_S$ $\in \{0.6\,,1.0\}$ fm. The matrix 
elements were found to be stable at the few percent level. 

We also analyzed the sensitivity of 
the GT-AA matrix elements to variation in 
the regulator function $F(r)$ defined as
\begin{equation}
\label{eq:fcut}
 F(r)=1 -\frac{1}{(r/R_L)^6 e^{[2(r-R_L)/R_L]}+1} \ ,
\end{equation}
for values of $R_L$ $\in \{0.6\,,0.8\}$ fm. We found
a variation of $\lesssim 17\%$ in the calculated
isospin-changing matrix elements of $A=$8--12 decays, 
a somewhat large variation which arises from a delicate cancellation
in the associated GT-AA transition densities (see Sec.~\ref{sec:results}
for explanation).
A detailed study focused on the cutoff dependence
is beyond the scope of this work, and
in what follows we report the matrix elements obtained without the regulator function given above. 
It would indeed be interesting to reanalyze these systems using 
different nuclear Hamiltonians. This  would  allow one to assess the sensitivity
to short-distance dynamics and to associate a model dependence
uncertainty to the calculations. In particular, Quantum
Monte Carlo calculations based on chiral two- and three-body 
potentials are now feasible~\cite{Lynn14,NV,Piarulli17}, 
which opens up the possibility of systematically and consistently
studying the sensitivity to 
cutoff variations in both the nuclear Hamiltonian and 
$0\nu\beta\beta$-decay potentials. Work along these lines is in progress.

\section{Results}
\label{sec:results}

\begin{figure}[bt]
\centering
\includegraphics[width=3.5in]{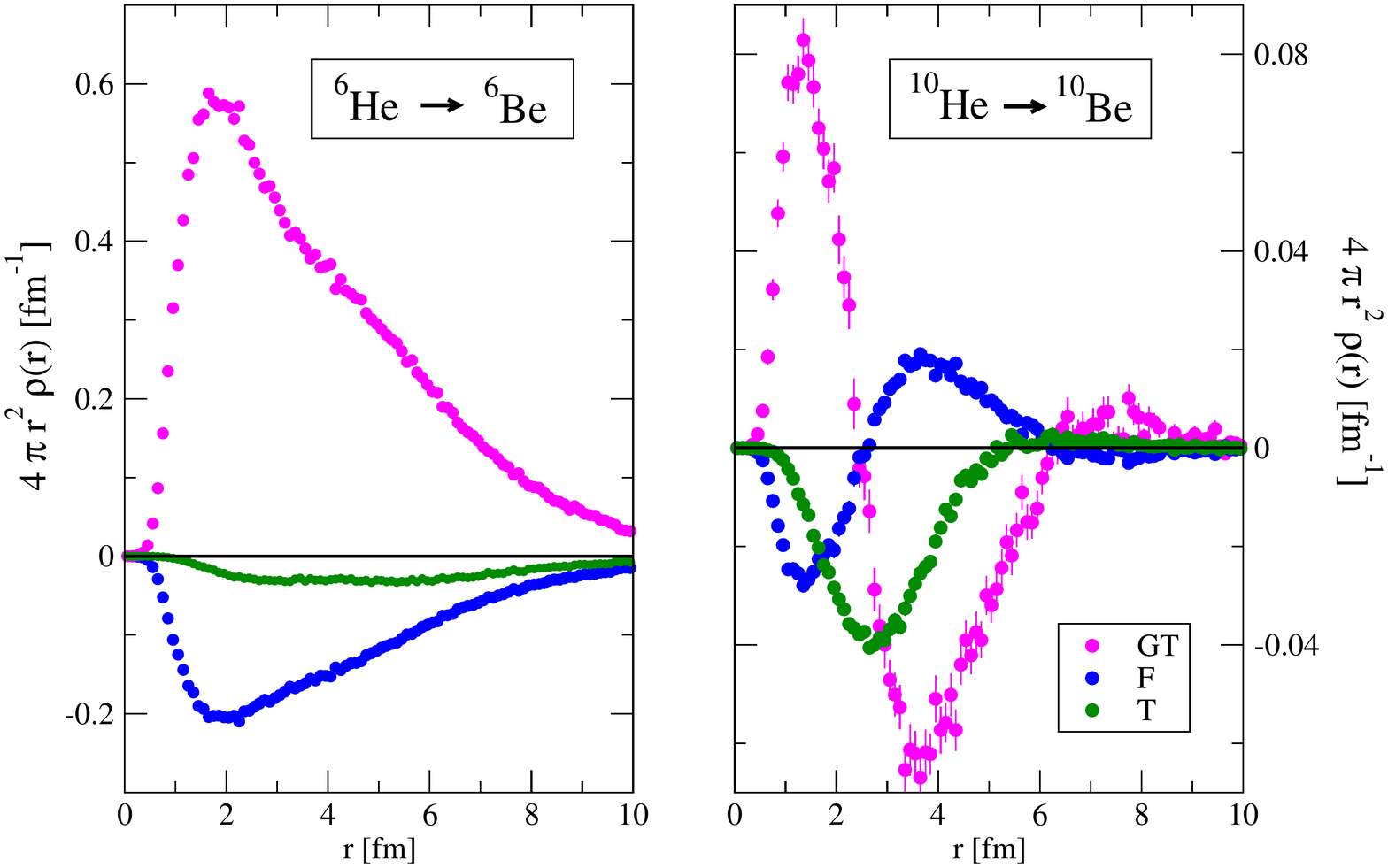} 
 \caption{VMC calculations of the transition densities associated with the F, GT, 
          and T operators---$\sum_{a<b}(\tau_a^+\tau_b^+)$,
          $\sum_{a<b}(\boldsigma_a \cdot \boldsigma_b\,\tau_a^+\tau_b^+)$, and
          $\sum_{a<b}(S_{ab}\,\tau_a^+\tau_b^+)$, respectively---for the $^{6}$He$\rightarrow^{6}$Be (left panel)
        and $^{10}$He$\rightarrow^{10}$Be decays (right panel).}
\label{fig:a610dens}
\end{figure}

Before proceeding to the discussion of the results, we emphasize
that we use the 
value of the 
axial coupling constant $g_A=1.2723 (23)$~\cite{Olive:2016xmw}.
In fact, recent GFMC studies on single-beta decay 
in $A\le10$ nuclei  based one the AV18+IL7 model adopted here, 
indicate that the ``$g_A$-problem''---that is 
the systematic over-prediction of 
single-beta Gamow-Teller matrix elements in 
simplified nuclear calculations---can be resolved 
by correlation effects in the nuclear wave functions~\cite{Pastore17}. 
These findings are limited to studies 
of matrix elements at zero  momentum transfer,
whereas the average momentum transfer
in $0\nu\beta\beta$-decay matrix elements 
is of the order of $\sim 100$ MeV~\cite{Engel:2016xgb}. It remains to be determined
how the ``$g_A$-problem'' propagates at intermediate
values of momentum transfer,
and whether the microscopic picture of the nucleus
based on the ``unquenched'' nucleonic weak couplings
successfully explains the data in this energy regime.
Progress in this direction would be facilitated 
by the acquisition of neutrino-nucleus scattering
data, which are scarce at moderated values of 
momentum transfer. 

In Tables~\ref{tb:2bme_standard} and \ref{tb:2bme}, we list the calculated $0\nu\beta\beta$-decay 
matrix elements in $^6$He, $^8$He, $^{10}$Be, $^{10}$He, and $^{12}$Be transitions.
We identify two classes of transitions, namely transitions in which the total isospin 
of the initial and final states remains unchanged, {\it i.e.}, $\Delta T=|T_i-T_f|=0$, 
and those  in which the total isospin changes by two units, {\it i.e.}, $\Delta T =2$.
The former involves isobaric analog states, which is never the case in
nuclear transitions considered for the actual experiments. It is nevertheless 
interesting to study these systems with the goal of benchmarking
different nuclear models and/or computational methods. 

Transition densities between isobaric analog states are
characterized by the lack of nodes:  this can be appreciated
in the left panel of Fig.~\ref{fig:a610dens} where we 
show results for the $^6$He$\rightarrow ^6$Be decay as 
a representative of this class. Once the VMC nuclear wave 
function for,  {\it e.g.} $^6$He, is determined,  then that of $^6$Be 
is obtained from it by swapping  protons and neutrons. As 
a result, the initial and final wave functions  
differ only in the third component of the isospin, 
while their radial and spin dependence is the same, implying
a maximum overlap between the two wave functions and the consequent
lack of nodes in the transition densities. In fact, evaluation of 
the $\sum_{a<b}\tau_a^+\,\tau_b^+$ operator in between these 
wave functions gives one, {\it i.e.}, the wave function normalization
(this is in case one neglects tiny contributions induced 
by the isoscalar Coulomb term~\cite{Pervin07}
which is different in the two isobaric analog nuclei 
due to their different number of protons). 
Similar considerations apply to the $A=10$ transitions
in this class. The $^8$He and $^8$Be$^\star$
excited state have the same spatial symmetry,
predominantly a $^1$S$_0$-[422], but with different
$T_z$ component. In fact, they both have an alpha-like core
with $S=T=0$, whereas the remaining two-nucleon 
pairs are two $^1$S$_0$-($nn$) dineutrons in $^8$He, and 
an equal mixture of two $^1$S$_0$-($np$) $T=1$ pairs,
one $^1$S$_0$-($nn$) dineutron and one $^1$S$_0$-($pp$) diproton
in $^8$Be. Again, there is no change in the spatial
symmetry of the initial and final states.

$\Delta T=2$ transitions are especially interesting
due to their direct correspondence to the 
experimental cases. As an example of this class, in the right 
panel of Fig.~\ref{fig:a610dens} we show the   
$^{10}$He$\rightarrow^{10}$Be transition densities associated with the F, GT, and 
T operators, namely $\sum_{a<b}(\tau_a^+\tau_b^+)$, $\sum_{a<b}(\boldsigma_a \cdot \boldsigma_b\,\tau_a^+\tau_b^+)$,
and $\sum_{a<b}(S_{ab}\,\tau_a^+\tau_b^+)$, respectively.
Here, the F and GT densities present a node
due to the orthogonality between the
dominant spatial symmetries of the initial [4222]=[$\alpha$,($nn$),($nn$),($nn$)]
and final [442]=[$\alpha$,$\alpha$,($nn$)] wave functions. Note
that integrating the F transition density (blue dots labeled with `F' in the figure) over $d{\bf r}$ 
 gives zero.  
Similarly, a node is found in the F and GT 
densities associated with the $A=8$ and $12$ 
transitions in this class. In particular,
the node is due to the orthogonality 
between the  dominant spatial symmetries
of the initial [422]=[$\alpha$,($nn$),($nn$)] 
([4422]=[$\alpha$,$\alpha$,($nn$),($nn$)])
and final [44]=[$\alpha$,$\alpha$] ([444]=[$\alpha$,$\alpha$,$\alpha$]) 
states in the $^8$He$\rightarrow^8$Be ($^{12}$Be$\rightarrow^{12}$C) decay.
In the remainder of this section we will primarily focus
our attention on $\Delta T=2$ transitions in $A=10$ and $12$, and just
report the results obtained for the $A=8$ decay. In fact, $^8$Be
presents a unique and rich structure characterized by a strong two-$\alpha$ clusters
in both its ground state---that lies $\sim 0.1$ MeV above the threshold for
breakup into two $\alpha$'s---and first two rotational excited states
of two $\alpha$ particles rotating about
each other~\cite{Pastore:2014oda,Datar:2013pbd}. 
These features make this test case less appealing for comparisons with 
decays relevant from the experimental point of view.

\subsection{Light Majorana neutrino exchange}

\begin{figure}[tb]
\centering
 \includegraphics[width=3.5in]{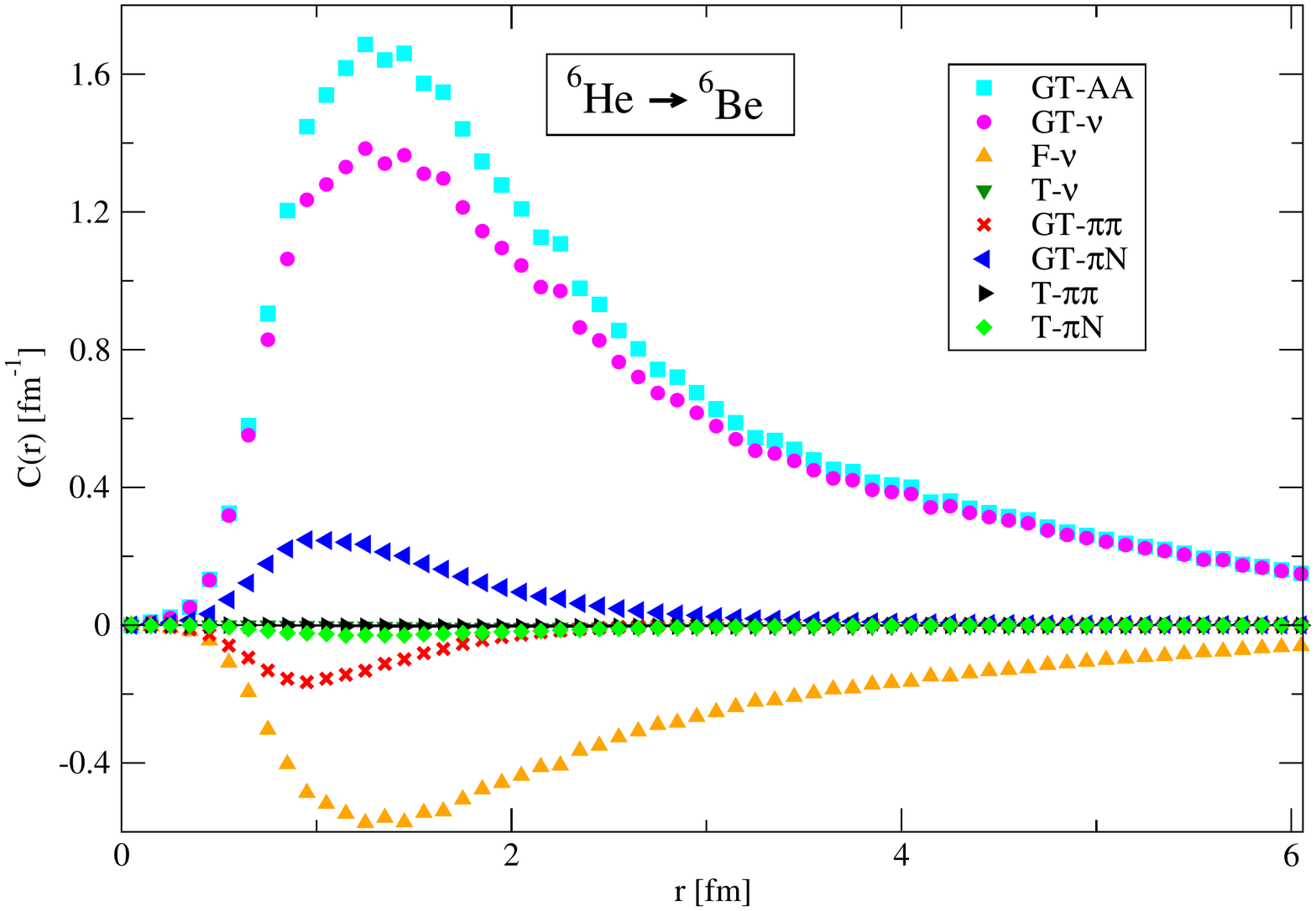} 
 \caption{VMC calculations of the transition distributions $C^{\alpha,\beta}(r)$ 
 defined in Eq.~(\ref{eq:densities}) for the $^{6}$He$\rightarrow^{6}$Be 
 decay.}
 \label{fig:he6r}
\end{figure}

In Table~\ref{tb:2bme_standard}, we report a breakdown 
of the tree-level light Majorana-neutrino exchange potentials
defined in Eqs.~\eqref{Vnu}--\eqref{smff}.
The first three rows show the results for transitions between isobaric analog states. In this case, 
the absence of the node implies that the F-$\nu$ and GT-AA contributions dominate the 
$0\nu\beta\beta$-potentials. 
The GT-AP and GT-PP components, which have pion-range,
steeply fall off for  $r \gtrsim 2$ fm, and give, respectively, a $\sim 20\%$ and $\sim 5\%$ correction to the GT-$\nu$ matrix element. 
This can be appreciated from Fig. \ref{fig:he6r}, which shows that for $r> 2$ fm the total GT distribution $C^{GT,\nu}$ is very well approximated 
by the AA component. The weak-magnetic term GT-MM, which is a N${}^2$LO correction in chiral EFT, is small, about $2\%$.
Fig. \ref{fig:he6r} also shows that the tensor matrix elements are negligible.

\begin{figure}[!b]
\centering
 \includegraphics[width=3.5in]{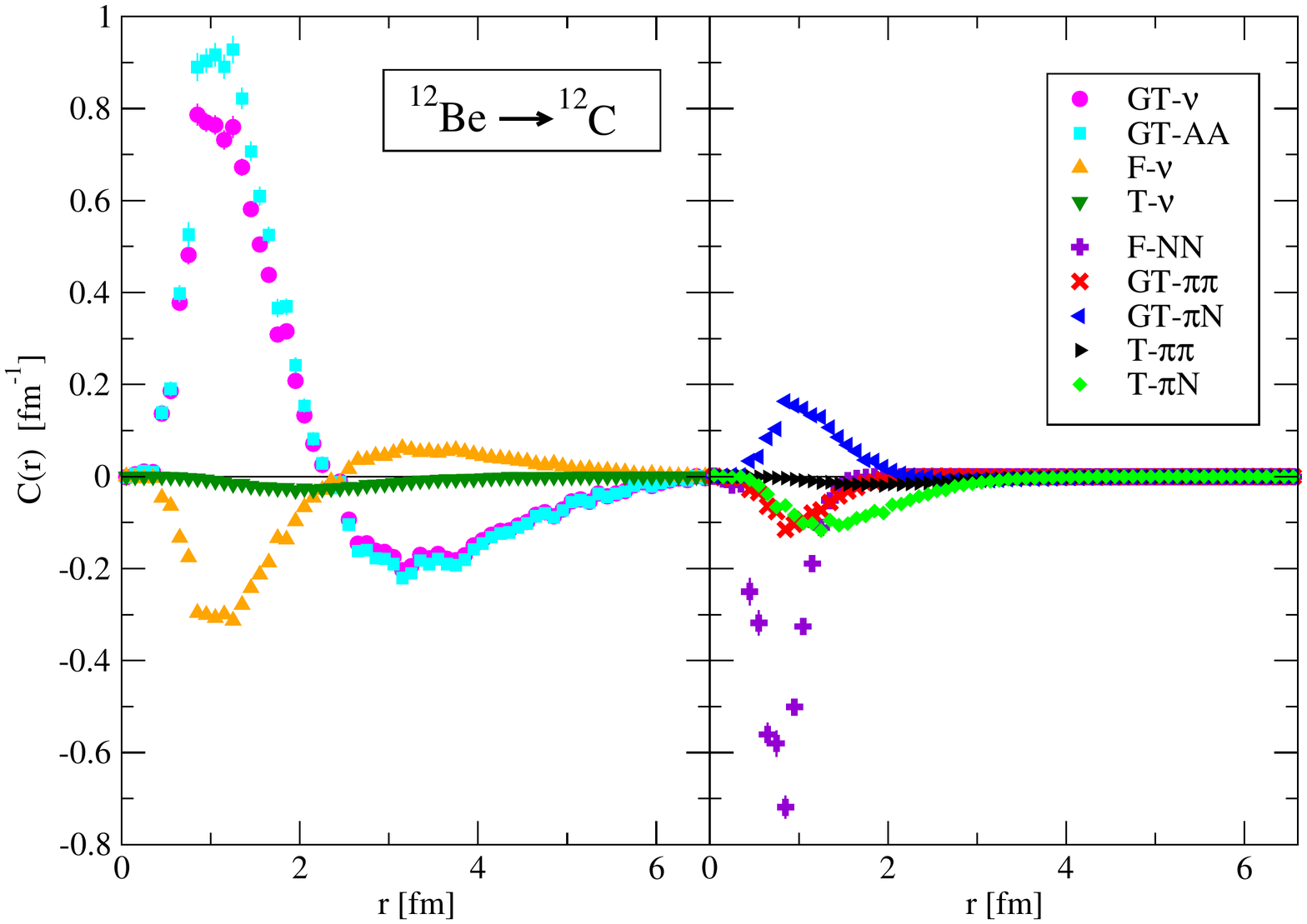} 
 \caption{VMC calculations of the transition distributions $C^{\alpha,\beta}(r)$ 
 defined in Eq.~(\ref{eq:densities}) for the $^{12}$Be$\rightarrow^{12}$C 
 decay.}
 \label{fig:be12r}
\end{figure}

\begin{table*}
 \caption{VMC calculations of the dimensionless matrix elements, defined in 
 Eq.~(\ref{MEdef}),   relevant for light Majorana-neutrino exchange.   
The first (second) three rows show the results for the $\Delta T=0$ ($\Delta T=2$) transitions 
(see text for explanation). For comparison, the bottom three rows show the results of \cite{Javier} 
for the heavy nuclei $^{48}$Ca, $^{76}$Ge, and $^{136}$Xe.
 VMC statistical errors (not reported in the table) are $\lesssim 2\%$.  
 }
 \setlength{\tabcolsep}{4.6pt}
  \begin{tabular}{c|c|c c c c c| c c c c |c}
\hline
\hline
$(T_i)\rightarrow (T_f)$ & 
\multicolumn{1}{c}{F}                 &
\multicolumn{5}{c}{GT}                  &
\multicolumn{5}{c}{T}                  \\
\cline{2-12}
\vspace{-0.1in}\\
                             &  \multicolumn{1}{c}{$\nu$}  &
    \multicolumn{1}{c}{$AA$} & \multicolumn{1}{c}{$AP$}   & \multicolumn{1}{c}{$PP$} &  \multicolumn{1}{c}{$MM$} & \multicolumn{1}{c}{$\nu$} & 
    \multicolumn{1}{c}{$AP$}          & \multicolumn{1}{c}{$PP$}& \multicolumn{1}{c}{$MM$} &\multicolumn{1}{c|}{$\nu$} & \multicolumn{1}{c}{$AA$}\\
\hline \hline  
{ $^6$He(1)$\rightarrow^6$Be(1)}          & -1.502  & 4.114  & -0.692  & 0.164 & 0.103 & 3.688 & -0.032    & 0.010 & -0.004& -0.025 & -0.099 \\
{ $^8$He(2)$\rightarrow^8$Be$^\star$(2)}  & -3.310 & 3.132  & -0.548  & 0.134 & 0.082 & 2.798 & -0.009    & 0.000 	   &   0.000   & -0.009 & -0.060\\
{ $^{10}$Be(1)$\rightarrow^{10}$C(1)}     & -1.898  & 4.326& -0.834  & 0.216 & 0.139 & 3.848& -0.097    & 0.032 &  -0.012   & -0.078 &-0.255 \\
 \hline \hline
{ $^8$He(2)$\rightarrow^8$Be(0)}       
                                          & -0.097  & 0.152& -0.117  & 0.042& 0.030& 0.108 & -0.026    & 0.010 & -0.004& -0.021 & -0.058 \\
{ $^{10}$He(3)$\rightarrow^{10}$Be(1)}  
                                          & -0.078  & 0.196  & -0.094  & 0.032 & 0.020& 0.156& -0.032    & 0.012 & -0.004& -0.026 & -0.074 \\
{ $^{12}$Be(2)$\rightarrow^{12}$C(0)}   
                                          & -0.192 & 0.500& -0.240  & 0.084 & 0.056& 0.400 & -0.066    & 0.024& -0.010& -0.052 & -0.142\\
\hline
\hline
{ $^{48}$Ca $\rightarrow^{48}$Ti}          & -0.25  &  1.08  & -0.38 & 0.13 & 0.10  & 0.93 & -0.08 &  0.03 & -0.01 & -0.06 & --\\
{ $^{76}$Ge $\rightarrow^{76}$Se}          & -0.59  &  3.15  & -0.94 & 0.30 & 0.22  & 2.73 & -0.01 &  0.00 &  0.00 & -0.01 & --\\
{ $^{136}$Xe $\rightarrow^{136}$Ba}        & -0.54  &  2.45  & -0.79 & 0.25 & 0.19  & 2.10 &  0.01 & -0.01 &  0.00 &  0.00 & --\\
\hline
\hline
\end{tabular}
\label{tb:2bme_standard}
 \end{table*}

The results for the $\Delta T=2$ transitions are shown in rows $4-6$ of Table~\ref{tb:2bme_standard}.
The most important feature of these transitions is the presence of the node, which causes the GT 
and F densities, illustrated in the right panel of Fig.\ \ref{fig:a610dens}, to change sign at about 2.5 fm.
As a result, there is a large cancellation for the F-$\nu$ and GT-AA matrix elements, 
which causes these NMEs to be significantly smaller than in the case of transitions involving isobaric analog 
states.   This is illustrated in the left panel of Fig.\ \ref{fig:be12r} for the 
$^{12}$Be$\rightarrow^{12}$C transition, where the region with $r > 2.5$ fm 
reduces the GT-AA matrix element by 50\%.
The same NMEs were compared in $\Delta T=2$ and $\Delta T=0$ transitions of heavier systems, such as Ca$\to$Ti, in Refs.\ \cite{Menendez:2015kxa,Menendez:2014ena}, where a similar suppression of the NMEs in $\Delta T=2$ transitions was found.
In contrast, the AP, PP and MM components, which are pion- and short-range contributions, 
are much less affected by this cancellation, and are therefore more important in 
the $\Delta T=2$ transitions. Both of these effects can also be seen from Table~\ref{tb:2bme_standard}.
For example, in the $^{10}$He$\rightarrow^{10}$Be transition the AP, PP and MM components are, 
respectively, $48\%$, $16\%$ and $10\%$ of the GT-AA,
and, while the GT-AA matrix element is 20 times smaller than 
in the $^6$He$\rightarrow^{6}$Be transition, the AP, PP and MM matrix elements are 
only about a factor of 5 smaller. 
Table \ref{tb:2bme_standard} also shows a partial cancellation between 
the GT-AP and GT-PP and GT-MM components, which is a common feature of 
both  $\Delta T =0$ and $\Delta T=2$ transitions. 
As a result we find that the GT-$\nu$ matrix element is always 
dominated by the GT-AA component.
In the case of transitions between isobaric analogues, 
the GT-AA matrix element is 90\% of the total  GT-$\nu$ contribution, 
while in $\Delta T = 2$ transitions, it is approximately $80\%$.
A similar effect is observed in calculations of heavier systems, 
such as $^{48}$Ca and $^{76}$Ge \cite{Menendez:2008jp,Barea:2009zza,Hyvarinen:2015bda,Barea:2015kwa,Javier}.

The absolute size of the NMEs shows sizable variations between different $\Delta T=2$ transitions.
In particular, the matrix elements increase by a factor of 2.5 between the $^{10}$He$\rightarrow^{10}$Be 
and $^{12}$Be$\rightarrow^{12}$C transitions. This can be appreciated from Fig. \ref{fig:10vs12q},
where we show the GT-$\nu$ and F-$\nu$ transition distributions in momentum space. While the shape of the distributions is
very similar in the two transitions, the peak is significantly larger in $^{12}$Be$\rightarrow^{12}$C. 
This effect
may be due, at least partially, to a large difference in the 
spatial extent of the relevant wave functions.
The $^{10}$He system is only a resonance, unstable against breakup into
$^8$He+$2n$ by about 1 MeV.
Here we have employed a pseudo-bound (with an exponentially falling 
density at long range) VMC wave function that is quite diffuse, with a 
proton (neutron) rms radius of 1.95 (3.66) fm.
The $^{10}$Be, $^{12}$Be, and $^{12}$C nuclei are all bound systems, with
VMC wave functions that have proton (neutron) rms radii of 2.32 (2.50) fm, 
2.43 (2.99) fm, and 2.48 (2.48) fm, respectively.
GFMC calculations change these radii by less than $ 5\%$.
Thus, for the $A=10$ decay, two neutrons with an rms radius of 3.66 fm
must be converted to two protons at an rms radius of 2.32 fm, indicating 
a small spatial overlap between the initial and final wave functions
and consequently relatively small matrix elements.
In comparison, the $A=12$ decay only requires a shift from 2.99 fm to
2.48 fm, which leads to a significantly larger spatial overlap, and larger 
matrix elements. This last transition in $A=12$ is possibly the test
case that is most like \NLDBD \ decays in nuclei of experimental interest.

As a comparison, in the last three rows of Table~\ref{tb:2bme_standard}
we show the shell model results for $^{48}$Ca, $^{76}$Ge and $^{136}$Xe \cite{Javier,Menendez:2008jp,Iwata:2016cxn}.
Other many-body methods differ by a factor of 2-3 \cite{Engel:2016xgb}. 
Although the absolute sizes of these NMEs are larger by a factor of a few than  those of the $\Delta T=2$ 
transitions calculated here, the relative factors between the different NMEs seem to 
agree fairly well (see also Table \ref{tb:ratios}), indicating
that the relative size of long- and short-distant physics is independent
of the particular nuclear systems considered.

It is interesting to note that the $R_A$ normalization factor introduced 
in Eqs.~(\ref{eq:opF})--(\ref{eq:opT}) can induce some misjudgment when 
comparing results from different nuclei.  In fact, if we multiply the NMEs by $1/R_A$
(with $R_8=2.40$ fm, $R_{10}=2.58$ fm, and $R_{12}=2.75$ fm) we find a  remarkably
good agreement between 
short- and pion-range potentials evaluated in $A=12$ and $A=48$ with $R_{48}=4.36$ fm 
(and, to a lesser extent, $A=76$ and $A=136$ with $R_{76}=5.08$ fm and $R_{136}=6.17$ fm) decays. 
This could be due to the fact that 
short-range  operators depend on the nuclear 
density which is roughly the same in all nuclei.

 \begin{table*}
 \centering
 \caption{VMC results for the dimensionless matrix elements, defined in Eq.~(\ref{MEdef}),  
 relevant for the contributions of the dimension-nine operators in Eq.~\eqref{eq:Ldim9}. For comparison, we 
 also show the total matrix elements for the light Majorana neutrino mechanism.
The first (second) three rows show the results for the $\Delta T=0$ ($\Delta T=2$) transitions (see text for explanation). For comparison, the bottom three rows show the results of \cite{Javier} for the heavy nuclei $^{48}$Ca, $^{76}$Ge, and $^{136}$Xe.
 VMC statistical errors (not reported in the table) are $\lesssim 2\%$. 
 }
 \setlength{\tabcolsep}{4.6pt}
  \begin{tabular}{c| c c |  c c c c | c c c}
\hline
\hline
$(T_i)\rightarrow (T_f)$ & 
\multicolumn{2}{c}{F}                 &
\multicolumn{4}{c}{GT}                  &
\multicolumn{3}{c}{T}                  \\
\cline{2-10}
\vspace{-0.1in}\\
                         &  \multicolumn{1}{c}{$\nu$}  & \multicolumn{1}{c}{$NN$}     &
          \multicolumn{1}{c}{$\nu$}   & \multicolumn{1}{c}{$\pi\pi$} &  \multicolumn{1}{c}{$\pi N$} & 
                                                                                                          \multicolumn{1}{c}{$NN$} &
    \multicolumn{1}{c}{$\nu$}          & \multicolumn{1}{c}{$\pi\pi$}& \multicolumn{1}{c}{$\pi N$} \\
\hline
{  $^6$He(1)$\rightarrow^6$Be(1) }         & -1.502  & -0.586   &3.688 & -0.160&0.354& 1.740& -0.025& -0.009 &  -0.040 \\      
{ $^8$He(2)$\rightarrow^8$Be$^\star$(2)}  & -3.310 & -0.532&2.798& -0.128&0.276& 1.414& -0.009 &0.000 & 0.015\\
{ $^{10}$Be(1)$\rightarrow^{10}$C(1)}     & -1.898& -0.876&3.848& -0.218&0.432& 2.588& -0.078& -0.032 &  -0.148 \\
\hline\hline
{ $^8$He(2)$\rightarrow^8$Be(0)}         
					 & -0.097  & -0.198         & 0.108& -0.044 &0.058&  0.596&-0.021 &-0.010 &-0.053 \\
{ $^{10}$He(3)$\rightarrow^{10}$Be(1)}  
					 & -0.078  & -0.134         & 0.156 & -0.032 &0.046& 0.402& -0.026& -0.012&-0.057\\
{ $^{12}$Be(2)$\rightarrow^{12}$C(0)}   
					  & -0.192  & -0.370& 0.400 & -0.084 &0.120&  1.106&-0.052 &-0.022 &-0.122\\
\hline
\hline
{ $^{48}$Ca $\rightarrow^{48}$Ti}          & -0.25  & -0.64    & 0.93& -0.12 & 0.18 & 2.11 & -0.060 & -0.026 & -0.153\\
{ $^{76}$Ge $\rightarrow^{76}$Se}          & -0.59  & -1.46    & 2.73& -0.31 & 0.49 & 4.87 & -0.010 & 0.00 & -0.026\\
{ $^{136}$Xe $\rightarrow^{136}$Ba}        & -0.54  & -1.28    & 2.1& -0.26 & 0.42 & 4.25 & -0.010 & 0.00 & 0.026\\
\hline
\hline
\end{tabular}
\label{tb:2bme}
\end{table*}

The last column of Table \ref{tb:2bme_standard} reports our results for the matrix element T-AA, which does not contribute in the case of light Majorana-neutrino exchange,
but it is relevant in the presence of right-handed charged-currents \cite{Doi:1985dx,Muto:1989cd,Cirigliano:2017djv}. This matrix element is not often computed in the literature,
and in Ref. \cite{Cirigliano:2017djv} bounds on the right-handed operator $C^{(6)}_{\rm VR}$ were obtained setting $M_{T, AA} = 0$. If we naively assume that the ratio between 
the GT-AA and T-AA matrix elements is the same in heavy and light nuclei, a T-AA matrix element of the size reported in Table \ref{tb:2bme_standard} would affect the bounds on $C^{(6)}_{\rm VR}$
at the 20\% level. 

The results discussed in this section, summarized in Table \ref{tb:2bme_standard}, deal mostly with NMEs involved in  light Majorana-neutrino exchange. However, as noted in Ref.\ \cite{Cirigliano:2017djv}, linear combinations of the same NMEs  determine additional long-range contributions to \NLDBD\ 
mediated by dimension-six and -seven LNV semileptonic operators, that are not proportional to $m_{\bt\bt}$.

\subsection{LNV from short-distance}

 \begin{table*}
 \caption{
 The Table shows the same matrix elements as Table \ref{tb:2bme},  relevant for dimension-nine contributions, 
 now normalized to the GT-${AA}$ (GT-$\pi N$) matrix element in the left (right) panel. 
 For comparison, the results of \cite{Javier,Hyvarinen:2015bda} for $^{48}$Ca, $^{76}$Ge and $^{136}$Xe are shown.
 }
 \renewcommand{\arraystretch}{1.1}
\begin{tabular}{c c| c c | c  c c c  }
\hline
\hline
$(T_i)\rightarrow (T_f)$  &  & 
\multicolumn{2}{c}{F}                 &
\multicolumn{4}{c}{GT} \\
\cline{3-8}
&\vspace{-0.1in}
\\
                       &  &  \multicolumn{1}{c}{$\nu$}   & \multicolumn{1}{c}{$NN$}   &
             \multicolumn{1}{c}{$AA$} &            
             \multicolumn{1}{c}{$\nu$} 
             &   \multicolumn{1}{c}{$\pi\pi$} &   \multicolumn{1}{c}{$\pi N$}\\
\hline                    
{$^8$He(2)$\rightarrow^8$Be(0)}      
					& & -0.63 &   -1.37   & 1  
					&  0.71 & -0.28 & 0.38\\
{$^{10}$He(3)$\rightarrow^{10}$Be(1)} %
					& & -0.39 &    -0.71   & 1
					&  0.79 & -0.16 & 0.23\\
{$^{12}$Be(2)$\rightarrow^{12}$C(0)}  
					& & -0.38 &    -0.77   & 1  &
					0.80 & -0.17 & 0.24\\
\hline                    
{$^{48}$Ca $\rightarrow^{48}$Ti} & \cite{Javier}                & -0.23 & -0.60 & 1  
& 0.86 & -0.11 & 0.17\\
\hline
{$^{76}$Ge $\rightarrow^{76}$Se} &\cite{Javier}                 & -0.19 & -0.46 & 1 
& 0.87 & -0.10 & 0.15\\
{\color{white}$^{76}$Ge $\rightarrow^{76}$Se}    &\cite{Hyvarinen:2015bda} 			       & -0.32 & -0.63 & 1 
& 0.84 & -0.12 & 0.19\\
\hline
{ $^{136}$Xe $\rightarrow^{136}$Ba} & \cite{Javier}                  & -0.22 & -0.52 & 1
 & 0.86 & -0.10 & 0.17\\
{\color{white}$^{136}$Xe $\rightarrow^{136}$Ba} & \cite{Hyvarinen:2015bda}   & -0.28 & -0.48 & 1
& 0.84 & -0.11 & 0.16\\
\hline	

\hline
\end{tabular}
\qquad
\begin{tabular}{ c | c c  }
\hline
\hline
  
\multicolumn{1}{c}{F}                 &
\multicolumn{2}{c}{GT} \\
\cline{1-3}
&\vspace{-0.1in}
\\
 \multicolumn{1}{c|}{$NN$}    & 
                                                                                           
                \multicolumn{1}{c}{$\pi\pi$} &   \multicolumn{1}{c}{$\pi N$}\\
\hline 
  
					        3.38  
					         & -0.76 &1\\

					         2.86   
					         &   -0.68 & 1\\
					         3.08   
					         & -0.70 & 1\\
\hline 
         3.55 
         &  -0.68 &1\\   
\hline
         2.97  
         &  -0.63 & 1\\
          3.34 
          &  -0.66 & 1\\

\hline
    3.06
    & -0.59 & 1\\
      3.03  
      &-0.68 & 1\\
\hline	

\hline
\end{tabular}
\label{tb:ratios}
\end{table*}

We now discuss the neutrino potentials induced by dimension-nine operators, 
which do not involve neutrino exchange, but are pion- or short-range.
Our results are summarized in Table \ref{tb:2bme}, where the first and middle three rows 
give the $\Delta T=0$ and $\Delta T=2$ transitions, respectively. For comparison, 
the bottom three rows give the results of Ref.~\cite{Javier} for the corresponding NMEs in 
heavier systems. 

By power counting, with the definitions in Eqs. \eqref{Vnu}--\eqref{smff} and \eqref{eq:dim9pot},
one would expect all the NMEs in Table \ref{tb:2bme} to be of similar size.
In the case of the $\Delta T = 0$ transitions, however,  
the lack of nodes  is responsible for the dominance of the GT-$\nu$ and F-$\nu$ NMEs over the other matrix elements listed in Table~\ref{tb:2bme}.
The GT-$\pi\pi$ and GT-$\pi N$ contributions
are, respectively, only $\sim 5\%$ and $\sim 10 \%$ of
the GT-$\nu$ matrix element. As these NMEs are proportional to GT-PP and 
GT-AP matrix elements, this is what we would expect from the results in Table \ref{tb:2bme_standard}. In Figs.~\ref{fig:he6r} and~\ref{fig:be12r}
we can see how the transition distributions associated with
the pion-exchange operators $\pi\pi$ and $\pi N$ start 
to die off at $\sim 1$ fm, which is expected since
the range of these operators is approximately set by $1/m_\pi\sim 1.4$ fm.
We also note that T-like 
operators are highly suppressed, as can be seen from the figures as well as from Table~\ref{tb:2bme}. 
This is a consequence of the fact that the tensor 
operator $S_{ab}$ vanishes in between $nn$-pairs 
in relative S-wave, which is the dominant 
two-nucleon component at short distances. 

For the $\Delta T=2$ class, 
we show in Fig.~\ref{fig:be12r}
the calculated distributions of the $^{12}$Be$\rightarrow ^{12}$C 
transition. Due to the characteristic node
in the GT transition densities and the 
ensuing cancellation, the GT-$\pi\pi$ (GT-$\pi N$) 
matrix element of this class is found to be as large
as $\sim 30\%$ ($\sim 40 \%$) of the GT-$\nu$ contribution
(see Table~\ref{tb:2bme}). This is (numerically) consistent with the results for the GT-PP 
and GT-AP matrix elements of Table \ref{tb:2bme_standard}. One can again see that the GT-$\pi\pi$ 
and GT-$\pi N$ distributions start to fall off around 1.1 fm, and that
the T-like operators are highly suppressed for the $\Delta T=2$ transitions as well.
From comparing the last six rows of Table \ref{tb:2bme} one can see that 
the absolute sizes of the matrix elements calculated here are smaller by a 
factor of a few than those calculated for heavier systems.
In Table \ref{tb:ratios} we show the F and GT matrix elements normalized to the 
GT-AA and GT-$\pi N$ components, including, for heavy system, results obtained with two many-body methods,
the shell-model \cite{Javier} and 
the quasiparticle random phase approximation \cite{Hyvarinen:2015bda}.
From the left panel we see that, in a given method, the relative importance of long-, pion- and short-range potentials is fairly constant, 
and the hierarchy of matrix elements is the same for heavy and light nuclei. 
For pion- and short-distance matrix elements, we observe an even better agreement.  
As illustrated in the right panel, after normalizing 
to GT-$\pi N$, the normalized short-range matrix elements of light and heavy nuclei, and of heavy nuclei computed with different methods,
are consistent at the  20\% level or better.

Finally, to obtain the short-range matrix elements GT-NN and F-NN we used the regularization 
of the delta function potential in Eq.\ \eqref{eq:cutoff}. If we instead regulate the divergence
by using a dipole form factor, either $g_{V}(\vec q\sq)$ or $g_{A}(\vec q\sq)$, the NMEs vary by no more than a few percent.

\subsection{Sensitivity to form factors and correlations}
\label{sect:ffandcorr}

We now turn our attention to the sensitivity of the 
matrix elements to variations in the nucleonic 
form factors as well as variations in the 
nuclear wave functions' correlations. To this end we study in more
detail the $\Delta T=2$ transition $^{10}$He$\rightarrow^{10}$Be 
and report our results in Table~\ref{tb:a10}. 
The findings discussed in this section in relation to 
the $A=10$ decay apply to the other $\Delta T=2$ transitions 
considered in the present  work as well. 

The neutrino potentials in Eqs.~\eqref{Vnu}--\eqref{smff} include the vector and 
axial form factors $g_V(\vec q^2)$ and $g_A(\vec q^2)$, whose momentum dependence is
an N${}^2$LO correction in chiral EFT. To study the impact of these form factors, we 
repeated the calculation of the NMEs setting $g_V(\vec q^2) =1$ and $g_A(\vec q^2)= g_A$.
We report the results for the $^{10}$He$\rightarrow^{10}$Be transition in the second 
row of Table \ref{tb:a10}. 
For the F-$\nu$ and GT-$\nu$ matrix elements the effect of turning off the axial 
and vector form factors is mild, resulting in at most a 10\% increase. 
For the T-AP and the T-PP components, this effect appears to be larger, $\sim$ 20\%-30\%.
In $\Delta T=2$ transitions the variation is magnified by the 
cancellations that affect the F and GT-AA matrix elements. For comparison, 
in $\Delta T=0$ transitions the effect of turning off
the momentum dependence of $g_{V,A}(\vec q^2)$ is less than 5\%.

\begin{figure}[bt]
\centering
 \includegraphics[width=3.5in]{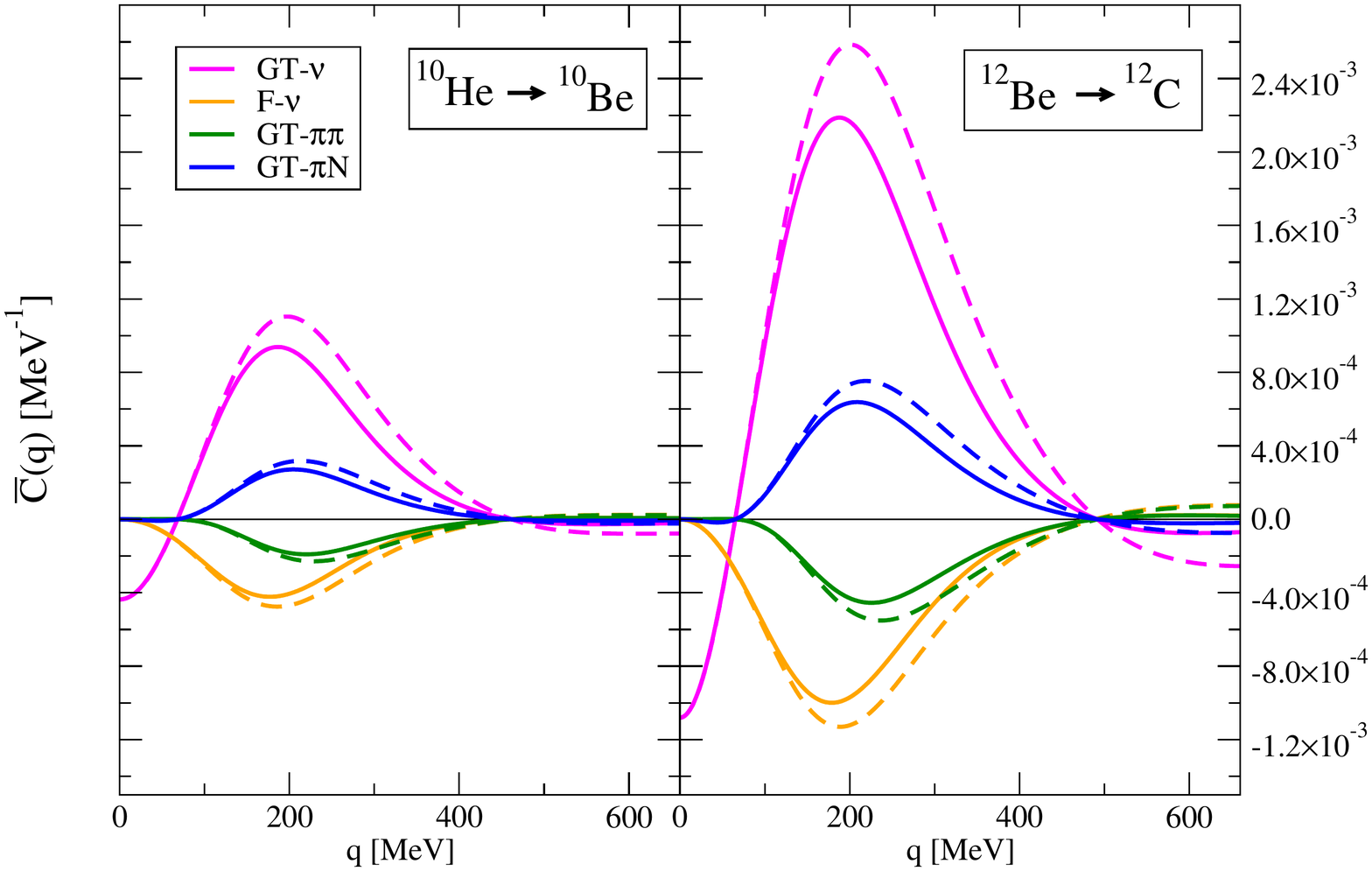} 
 \caption{The GT-$\nu$, F-$\nu$, GT-$\pi\pi$, and GT-$\pi N$ 
 distributions in momentum space for the $^{10}$He$\rightarrow^{10}$Be and $^{12}$Be$\rightarrow^{12}$C
 decays. Solid and dashed lines are obtained, respectively, 
with and without the inclusion of the momentum dependence in nucleonic form factors.
See text for explanation.}
\label{fig:10vs12q}
\end{figure}

\begin{figure}[bt]
\centering
 \includegraphics[width=3.5in]{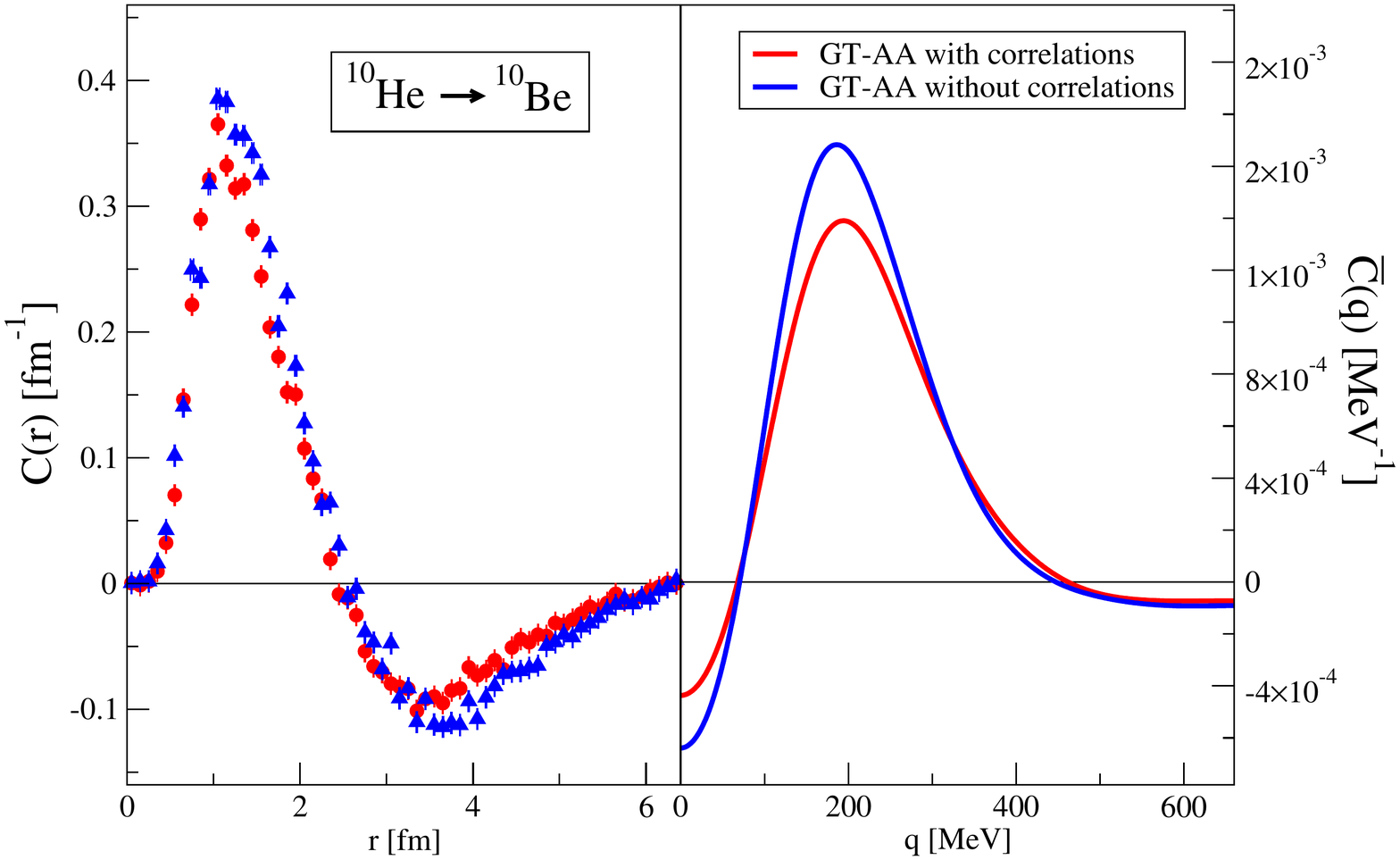} 
 \caption{
 The left (right) panel shows the GT-AA  distribution in $r$-space ($q$-space) 
 for the $^{10}$He$\rightarrow^{10}$Be  transition, with and 
 without ``one-pion-exchange-like'' correlations 
          in the nuclear wave functions. See text for explanation.  
 }
\label{fig:corr10}
\end{figure}

For the weak-magnetic contributions GT-MM, some care has to be taken when removing
the form factors. As evident from Eqs.\ \eqref{ourdef.d} and \eqref{ourdef.t}, 
in the absence of $g_V(\vec q^2)$, both $V_{GT, MM}$
and $V_{T, MM}$ are singular at $r\rightarrow 0$. To compute the 
GT-MM matrix element in the second line of Table \ref{tb:a10} we used the regularization of the delta 
function in Eq.~(\ref{eq:cutoff}), with $R=0.6$ fm. Varying $R$ between $0.6$ and $0.8$ 
fm does not have an appreciable effect on the result.  The good agreement for the values of GT-MM 
in the first and second line of Table \ref{tb:a10} indicates that the result does not strongly depend 
on the way the region of large $\vec q^2$ is regulated.  For the T-MM matrix element, the second line 
of Table \ref{tb:a10} is obtained by naively 
using the potential $V_{T, MM}(r)$ in Eq. \eqref{ourdef.t}.  
Here the divergence at $r=0$ does not spoil the
evaluation of the associated matrix element. Again this
is due to the fact that the tensor operator T ($S_{ab}$) 
gives zero on pairs in relative S-wave. In fact, 
the $\tau_a^+\tau_b^+$ is selecting out valence ($nn$) pairs
in the initial state. These are largely in a $^1S_0$ relative 
state, with some $^3P_0$ components which are however zero 
at short-range due to an angular momentum barrier.

While in Table \ref{tb:a10} we only report results for the impact of form factors on the light neutrino-exchange potentials, the same 
 features are shared by matrix elements of the $V_{\pi\pi}$ and  $V_{\pi N}$ potentials, as they are proportional to 
to the AP and PP components in \ref{tb:a10}. The same holds for the $V_{NN}$ potential, which is analogous to GT-MM. 
In particular, changing the regularization  
of the delta function potential from Eq.\ \eqref{eq:cutoff} to a dipole form factor, 
either $g_{V}(\vec q\sq)$ or $g_{A}(\vec q\sq)$
has little effect on the F-NN and GT-NN matrix elements.

The impact of the axial and vector form factors on the $^{10}$He$\rightarrow^{10}$Be and 
$^{12}$Be$\rightarrow^{10}$C transitions is illustrated in Fig.\ \ref{fig:10vs12q}.
The solid and dashed lines denote the distributions $\bar{C}(q)$ defined in Eq.\ \eqref{eq:densities},
with and without the dipole form factors for $g_{V,A}(\vec q^2)$.
We see that the dipole form factors start to have an effect at around $q \sim 200$ MeV, 
and cut off the distributions for $q \gtrsim 500$ MeV. 
The effect is similar for the F-$\nu$ and GT-$\nu$, which are mostly long-distance, 
and the pion-range GT-$\pi\pi$ and GT-$\pi N$ matrix elements, which are induced by heavy LNV new physics.       

In the third row of Table~\ref{tb:a10}, we report results obtained by regulating the
matrix elements with the $F(r)$ function defined in Eq.~(\ref{eq:fcut}) with $R_L=0.7$ fm.
We studied the sensitivity of our results with respect to variation of $R_L\in\{0.6,0.8\}$ fm
and found that the most affected matrix elements are those characterized by the presence
of the node. For example, by comparing the second and the third rows in the table we can see
that GT-$\nu$ and F-$\nu$ undergo a $\sim 18\%$ and $\sim 13\%$ variation, respectively,
whereas T-$\nu$ is essentially unaffected by the regulator function. This is because the
T-like operators are already zero at short-distances.

{\centering
 \begin{table*}
 \caption{VMC calculations of the dimensionless matrix elements relevant for
 light Majorana-neutrino exchange, defined in Eqs.~(\ref{ourdef.a})--(\ref{ourdef.c}),
 for the $^{10}$He$\rightarrow^{10}$Be transition. 
 The first row repeats the results of Table \ref{tb:2bme_standard}, 
 which include both the form factors and correlations. The results reported in the second row neglect the momentum dependence
 in the axial, vector and pseudoscalar nucleonic form factors. Results in the third row are obtained
 including the regulator given in Eq.~(\ref{eq:fcut}). Results in the forth row are obtained
 turning off the ``one-pion-exchange-like'' correlations in the nuclear wave functions (see text for explanation).
 VMC statistical errors (not reported in the table) are $\lesssim 2\%$.  
 }
 \setlength{\tabcolsep}{4.6pt}
  \begin{tabular}{c|c|c c c c c| c c cc}
\hline
\hline
$(T_i)\rightarrow (T_f)$ & 
\multicolumn{1}{c}{F}                 &
\multicolumn{5}{c}{GT}                  &
\multicolumn{4}{c}{T}                  \\
\cline{2-11}
\vspace{-0.1in}
\\
                        &  \multicolumn{1}{c}{$\nu$}  &
    \multicolumn{1}{c}{$AA$} & \multicolumn{1}{c}{$AP$}   & \multicolumn{1}{c}{$PP$} &  \multicolumn{1}{c}{$MM$} & \multicolumn{1}{c}{$\nu$} & 
    \multicolumn{1}{c}{$AP$}          & \multicolumn{1}{c}{$PP$}& \multicolumn{1}{c}{$MM$} &\multicolumn{1}{c}{$\nu$} \\
\hline
{ $^{10}$He(3)$\rightarrow^{10}$Be(1)}   		& -0.078  & 0.196  & -0.094  & 0.032 & 0.020 & 0.156 & -0.032    & 0.012 & -0.004& -0.026 \\ 
  no form factors                                       & -0.088  & 0.218  & -0.098  & 0.034 & 0.020 & 0.172 & -0.042    & 0.016 & -0.006     & -0.032\\
  $F(r)$, $R_L = 0.7$ fm                           & -0.076  & 0.180  & -0.086  & 0.028 & 0.013 & 0.141 & -0.041    & 0.015 & -0.006     & -0.033     \\    
no correlations                                  & -0.086  & 0.222  & -0.106  & 0.036 & 0.022 & 0.172& -0.004    & 0.002 &   0.000 & -0.004 \\
\hline
\hline
\end{tabular}
\label{tb:a10}
 \end{table*} }

Finally, in the forth row of Table~\ref{tb:a10} 
we report results obtained by artificially turning off 
the ``one-pion-exchange-like'' correlation operators
in the nuclear wave functions as  discussed in Sec.~\ref{sec:vmc}. 
Turning the correlations off has a dramatic effect on the 
tensor matrix elements, which become  statistically equal to zero.
The GT-$\nu$ and F-$\nu$ magnitudes 
increase by  $\sim 10\%$ with respect to the correlated
results given in the first row of the table. 
The  effect of the ``one-pion-exchange-like'' correlations
is represented in Fig.~\ref{fig:corr10}, where
the blue triangles (solid line) in the left (right)
panel represent the $r$-space ($q$-space) GT-AA transition
distribution obtained by turning off the correlations
to be compared with the red dots (solid line) 
obtained with the correlated wave function.

In closing this section, we 
reiterate that   \NLDBD \ matrix 
elements involve on average values of momentum 
transfer $q$ of the order of hundreds of MeVs.
This can be seen, for example, in Fig.~\ref{fig:10vs12q} 
where the momentum distributions in both the $A=10$ 
and $12$ decays peak at $\sim 200$ MeV.

\subsection{Light neutrino exchange beyond leading order}
\label{Sect:Vnu2}
Beyond leading order, several new contributions to light Majorana-neutrino exchange arise. At N$^2$LO in the Weinberg counting, these consist of corrections to the single-nucleon currents as well as genuine two-body effects that cannot be absorbed by the one-body weak currents \cite{Cirigliano:2017tvr}.  The second effect is induced by loop diagrams involving the neutrino, as well as counterterms that appear at the same order. 
The corrections to the one-body currents are often included in  the \NLDBD\ literature through the form factors in Eq.\ \eqref{eq:FF}, while the two-body contributions have so far not been implemented in nuclear calculations. Here we investigate the impact of this second type of corrections, which appears at the same order as the effect of the form factors discussed in Section~\ref{sect:ffandcorr}.

The N$^2$LO correction to the neutrino-exchange potential of Eq.\ \eqref{eq:vnu} was derived in Ref.\ \cite{Cirigliano:2017tvr} and can be written as 
\bea \label{eq:N2LOpot}
V_{\nu,2} = \tau_a^+\tau_b^+ \left(V_{VV}^{(a,b)}+V_{AA}^{(a,b)}+V_{CT}^{(a,b)}+ V_{\rm us}^{(a,b)} \ln\frac{m_\pi\sq}{\mu_{\rm us}\sq}\right),\nn\\
\eea
where $V_{VV}^{(a,b)}$  ($V_{AA}^{(a,b)}$) arises from loops with two insertions of the vector (axial) current, $ V_{\rm us}$ is generated by loops involving ultrasoft neutrinos, and $V_{CT}^{(a,b)}$ captures the counterterm contributions.
The latter term involves three counterterms which absorb the renormalization scale ($\mu$) dependence of  divergent loop diagrams. We write these pieces as follows
\footnote{With these definitions, $V_{VV,AA}$ and $ V_{\rm us}$ correspond to $\mathcal V_{VV,AA}$ and $\tilde {\mathcal V}_{AA}$ of Ref.\ \cite{Cirigliano:2017tvr} with $L_\pi=0$, while  $ V_{CT}$ includes  $\mathcal V_{CT}$ as well as the $L_\pi$ pieces of $\mathcal V_{VV,AA}$. We neglected the contribution of the contact interaction, $C_T$, everywhere.},
\bea
V_{CT}^{(a,b)} &=&\left(\frac{5}{6}g_\nu^{\pi\pi}+3L_\pi\right)V_{CT,\pi \pi}^{(a,b)}\nn\\
&&+ \left(g_\nu^{\pi N}+(1-g_A\sq)L_\pi\right)V_{CT,\pi N}^{(a,b)}\nn\\
&&+ \left(g_\nu^{NN}+\frac{3}{8}(1-g_A\sq)\sq L_\pi\right)V_{CT,NN}^{(a,b)}\,\,,
\eea
where $L_\pi = \ln \frac{\mu\sq}{m_\pi\sq}$ and $g_\nu^{\pi\pi}$,$g_\nu^{\pi N}$, and $g_\nu^{ NN}$ are the counterterms. 

It should be noted that the potential in Eq.\ \eqref{eq:N2LOpot} does not capture the complete N$^2$LO correction. Firstly, 
the loops involving ultrasoft neutrinos (captured by $ V_{\rm us}$) are divergent and induce the dependence on the renormalization scale $\mu_{\rm us}$ in Eq.\ \eqref{eq:N2LOpot}. This $\mu_{\rm us}$ dependence is canceled by ultrasoft contributions to the \NLDBD\ amplitude. 
However, the calculation of these contributions requires knowledge of the intermediate states \cite{Cirigliano:2017tvr} and is beyond the scope of the current work. 
Secondly, although $g_\nu^{\pi\pi}$ can be estimated through a connection to electromagnetic corrections to $\pi\pi$  interactions~\cite{Ananthanarayan:2004qk}, 
leading to~\cite{Cirigliano:2017tvr} $g_\nu^{\pi \pi} (\mu = m_\rho) = - 7.6$, 
the counterterms $g_\nu^{\pi N}$ and $g_\nu^{ NN}$ are currently unknown. Without these missing pieces we do not have full control over the complete N$^2$LO correction.
Nevertheless, a rough estimate of the size of the  counterterm and the ultrasoft contributions can be obtained by varying the renormalization scales, $\mu$ and $\mu_{\rm us}$, respectively, 
such that the  logarithms change by O(1)  (this corresponds to Naive Dimensional Analysis (NDA)).

With the above caveats in mind, we find in the case of the $^{10}$He$\to ^{10}$Be transition
\bea
\frac{M_{VV}}{M_\nu} &=& 7.1\Ex{-3} ,\quad \frac{M_{AA}}{M_\nu} = -7.9\Ex{-2},\nn\\
 \frac{M_{CT,\pi\pi}}{M_\nu} &=& 8.5\Ex{-3},\quad \frac{M_{CT,\pi N}}{M_\nu} = -3.8\Ex{-3}, \nn\\
 \frac{M_{CT,NN}}{M_\nu} &=& 1.4\Ex{-2}, \quad\frac{ M_{\rm us}}{M_\nu} = -2.4\Ex{-2} \,\, ,
 \label{eq:N2LOme}
\eea
where $M_\nu$ denotes the matrix element of the potential in Eq.\ \eqref{Vnu}, $M_\nu = -M_{F,\nu} + g_A\sq (M_{GT,\nu}+ M_{T,\nu})$ which can be read from Table \ref{tb:2bme_standard}. For the $^{10}$He$\to ^{10}$Be transition, one has $M_\nu  \simeq 0.29$. It should be noted that the potential in Eq.\ \eqref{eq:N2LOpot} has a divergence for $q\to \infty$ (or $r\to 0$), making it rather sensitive to the way short-distance scales are regulated. Here we naively regulated this divergence by multiplying all terms by $g_A\sq(\vec q\sq)/g_A\sq$. 

The sizes of the different pieces in Eq.\ \eqref{eq:N2LOme} vary from the sub-percent level to O(10\%)  of the LO matrix element, $M_\nu$, which is consistent with the expected size of  N$^2$LO corrections. As a result, some of the larger terms in Eq.\ \eqref{eq:N2LOme} are of the same order of magnitude as the effects of including the form factors. 
NDA estimates of the counterterms do not alter this conclusion. 
However, one should note that the NDA scaling of $g_\nu^{NN}$ is far from obvious in the context of chiral EFT. 
As discussed in Ref.~\cite{Cirigliano:2017tvr},  further work to determine the scaling of $g_\nu^{NN}$ and its possible enhancement is needed.  

\section{Conclusion}
\label{sect:concl}

The nuclear {\it ab initio} approach aims at describing 
the widest range of nuclear properties in terms of interactions 
occurring between nucleons inside the nucleus. In this microscopic
picture, nucleons interact with each other via two- and three-body 
interactions, and with external electroweak probes via couplings
to individual nucleons and to nucleon-pairs. {\it Albeit} limited to
light nuclei ($A\le12$), Quantum Monte Carlo calculations based on the AV18 two-body 
and IL7 three-body interactions successfully explain available 
experimental data in a broad energy range, 
from the keV regime relevant to astrophysics studies to
the GeV regime where short-range correlations become 
predominant~\cite{Carlson15,Carlson98,review2014}.
These studies yield a rather complex picture of the nucleus
with many-body correlations in both the nuclear 
wave functions and electroweak currents playing
an important role in reaching agreement with the data. 

In this work, we used the {\it ab initio}  approach 
supported by the computationally accurate Quantum Monte
Carlo methods to study \NLDBD \ matrix elements 
in $A=6$--$12$ nuclei. While these systems are not
relevant from the experimental point of view, they 
are nevertheless interesting and provide us with an 
extremely useful set of test cases.
In fact, the  \NLDBD \ rate depends on matrix elements 
that are not experimentally accessible and need to 
be estimated theoretically.
At present, the calculated nuclear matrix elements
of experimental interest ($A\ge48$) have large theoretical 
uncertainties which  complicate the  interpretation of any future \NLDBD \ observation or lack thereof.
The uncertainties on the calculated
matrix elements are primarily attributable to the fact
that for larger nuclear systems, in order 
for the calculations to be computationally feasible, 
one has to (drastically) approximate the 
{\it ab initio} framework, by, {\it e.g.}, leaving
out correlations and/or truncate the model space.

It is in this context that this study on  \NLDBD \ in light nuclei  finds  its relevance. 
For a start, we provided a set of VMC calculations 
that can be used for benchmarking purposes. 
We have presented results for the nuclear 
matrix elements relevant for the light  Majorana-neutrino exchange mechanism 
(Table~\ref{tb:2bme_standard})  
as well as for TeV-scale mechanisms of lepton-number violation (Table~\ref{tb:2bme}), 
and we have studied their relative size (see Table~\ref{tb:ratios}). 

Our results  for the $\Delta T = 2$ transitions  show the following features:
(i) 
The  matrix elements  for $A=10,12$ are between an order of magnitude and 
a factor of two smaller compared to shell model results for systems with $A=48,76,136$. 
The  bulk of this difference can be attributed to the normalization factor $R_A$ entering 
Eqs.~(\ref{eq:opF})--(\ref{eq:opT}). 
(ii) 
The difference  in the $A=10$ and $A=12$ matrix elements is correlated 
with the height of the peaks in their associated transition densities 
(see Fig.~\ref{fig:10vs12q}) and it is due to the different spatial overlaps 
between an initial diffuse neutron distribution and a final compact proton 
distribution in the case of the $A=10$ transition, and between two compact 
initial neutron and final proton distributions in the $A=12$ transition. 
(iii)  
As illustrated in Table~\ref{tb:ratios},   
the ratios of different matrix elements to the dominant Gamow-Teller one 
(GT-AA)  are, in a given method,  roughly  independent of $A$.  
We find that for  $A=10,12$, the ratios agree at the 5\% level,  
while  for  $A=48,76,136$  they agree at the 15\% level or better, 
and are consistent with the $A=10, 12$  results at the 30\% level.
However, if we normalize the GT-like 
matrix elements by a short-range contribution, {\it e.g.}, 
GT-$\pi N$, then the normalized short-range matrix 
elements are consistent  at the $\sim 20\%$ level or better 
in all the considered nuclear transitions.

Our results will help  the community assess the adequacy of the 
various  methods used to 
estimate \NLDBD \ matrix elements,
and identify the key dynamical features
that need to be retained in more approximate
many-body computational methods.
This is especially relevant for benchmarking
 those methods that can be extended 
to the heavier systems of experimental interest. 
In this spirit, 
we have studied the  effect 
of artificially turning off correlations in 
the VMC nuclear wave functions, 
finding a $\sim 10 \%$ increase in the
calculated nuclear matrix elements for the 
light Majorana neutrino exchange mechanism. 
In previous studies, we found that turning off correlations---as
described in  Section~\ref{sec:vmc}---and keeping only the dominant
component in the VMC w.f.'s leads to a $\sim 15\%$ ($\sim 30\%$)
increase in the calculated single beta decay matrix elements
of $A=6$--$7$ ($A=10$) transitions, with respect
to the fully correlated results that are in agreement
with the data at the $\leq2\%$ ($\sim 10\%$) level~\cite{Pastore17}.
This corresponds to having to ``quench'' $g_A$ by $q\sim 0.85$ ($q\sim 0.70$)
 in $A=6$--$7$ ($A=10$) single beta decays. 
This is a somewhat larger effect than what we have found here 
for the calculated \NLDBD \ matrix elements. For example,
in the $A=10$, $\Delta T=2$ \NLDBD \ transition we find a $\sim 25\%$ variation  
in the calculated matrix elements when we use the `uncorrelated' wave functions,
which corresponds to a $g_A$ ``quenching'' of $\sim 0.90$.
These findings may indicate that the $g_A$ ``quenching''
required in calculations based on more approximated nuclear models
(for  $A>12$ nuclei) is larger in single beta decay than in \NLDBD.

Within the VMC approach,  we have also  explored the impact of using different forms for the 
transition  operators  mediating \NLDBD\ -- another potential source of uncertainty 
in the matrix elements of physical interest. 
In particular, for the light Majorana-neutrino exchange mechanism, 
following the chiral EFT approach of Ref.~\cite{Cirigliano:2017tvr} 
we have estimated  the impact of N$^2$LO corrections (in the Weinberg power counting)
on the $^{10}$He$\rightarrow^{10}$Be transition. 
The ``factorizable" N$^2$LO effects captured by nucleon form factors impact the matrix elements at the 10\% level (see Table~\ref{tb:a10}). 
The non-factorizable genuinely two-body effects are discussed in Section~\ref{Sect:Vnu2}.  
While we do not have yet full control over the N$^2$LO amplitude (counterterms and ultrasoft contributions are not yet known), 
our results  suggest that the non-factorizable effects may lead to O(10\%) corrections, consistently with the expectations 
of the chiral power counting. 
Counterterms of the size implied by naive dimensional analysis would not change this conclusion. 
One should keep in mind, however,  that the NDA scaling of the four-nucleon coupling  $g_\nu^{NN}$ 
cannot  be taken for granted~\cite{Cirigliano:2017tvr},  
and further work to check the consistency of Weinberg power counting for \NLDBD\ and to determine the scaling of $g_\nu^{NN}$ is needed.  
In a similar vein, future work should focus on a more consistent chiral EFT approach, in which the nuclear 
wave functions are determined from a chiral potential.

\section{Acknowledgments}
We would like to thank Javier Men\'endez
for useful discussions at various stages 
of this work
and for providing us with updated shell-model nuclear matrix
elements before publication. 
We thank the  Institute for Nuclear Theory at the University of Washington for its hospitality and the Department of Energy 
for partial support during  the  program INT-17-2a,   during which  this work was initiated.
The work of S.P., J.C., and~R.B.W.~has been supported by the Nuclear
Computational Low-Energy Initiative (NUCLEI) SciDAC project. This research
is also supported by the U.S.~Department of Energy, Office of Science,  Office of
Nuclear Physics, under contracts 
DE-AC02-06CH11357 (R.B.W.), and DE-AC52-06NA25396 and 
Los Alamos LDRD program (J.C., V.C., E.M.). W.D.  acknowledges  support by the Dutch Organization for Scientific Research (NWO) 
through a RUBICON  grant. 
Computational resources have been provided by Los Alamos Open
Supercomputing, and Argonne's Laboratory Computing Resource Center.

\appendix

\section{Neutrino potentials in coordinate space}\label{App1}

Neglecting the momentum dependence of the axial and vector form factors, the potentials in coordinate space read 
\bea
V_\nu  &=&  m_\pi \tau_a^+ \tau_b^+  
\Big( 
\mathbf{1} \times \mathbf{1}  \ V^\nu_{F}(z)  
\nonumber \\
&-&   g_A^2 \,    \boldsigma_a \cdot \boldsigma_b  \, V^\nu_{GT}(z) 
\ -\ 
g_A^2    \, S_{ab} \,  V^\nu_{T}(z) \Big) \ ,
\nonumber \\
V_{\pi \pi} &=& -  m_\pi \tau_a^+ \tau_b^+   \, 
\left(   \boldsigma_a \cdot \boldsigma_b   \,   V_{GT, \pi \pi}(z)    
\ + \  S_{ab}  \,  V_{T,\pi\pi}(z)  \right)~, 
\nonumber \\
V_{\pi N} &=& -  m_\pi \tau_a^+ \tau_b^+   \, 
\left(   \boldsigma_a \cdot \boldsigma_b   \,   V_{GT, \pi N}(z)    
\ + \  S_{ab}  \,  V_{T,\, \pi N}(z)  \right)~, \nonumber \\
V_{NN} &= &  m_\pi\, \tau_a^+ \tau_b^+\,  V_{F,\, NN}(z) \ , 
\eea
where $S_{ab} (\hat{r}) \equiv  3\,\boldsigma_a \cdot  \hat{ \vec r} \, \boldsigma_b \cdot  \hat{ \vec r} - \boldsigma_a\cdot \boldsigma_b$, 
and we have introduced $z = r m_\pi$, with $r$ indicating the distance between
particles $a$ and $b$.  
The light Majorana neutrino exchange potentials $V_{F}^{\nu}$, $V_{GT}^{\nu}$ and $V_{T}^{\nu}$ are   
\begin{eqnarray} \label{ourdef.a}
V_{F,\, \nu}(z) &=& \frac{1}{4\pi z}, \\
V_{GT,\, \nu}(z) &=& V_{GT, AA}(z)  + V_{GT, AP}(z) \nonumber \\
&+& V_{GT, PP}(z) + V_{GT, MM}(z) \ ,\label{ourdef.b}\\ 
V_{T,\, \nu}(z)  &=& V_{T, AP}(z)  + V_{T, PP}(z)  + V_{T, MM}(z)\ , \label{ourdef.c}
\end{eqnarray} 
where the GT functions are given by 
\begin{eqnarray}\label{ourdef.d}
&& V_{GT, AA}(z)  = \frac{1}{4\pi z}\ , \quad  V_{GT, AP}(z)  = -  \frac{e^{-z}}{6\pi z}\ ,\nonumber \\
&& V_{GT, PP}(z)  = - \frac{e^{-z} (z-2)}{24\pi z} \ , \nonumber \\
&& V_{GT, MM}(z)  = \frac{ (1 + \kappa_1)^2 \, m_\pi^2}{6 g_A^2 m_N^2} \delta^{(3)} (m_\pi \vec r) \ .
\end{eqnarray}
The tensor functions are 
\begin{eqnarray}\label{ourdef.t}
&& V_{T, AP}(z)  = \frac{1}{4\pi z^3} \left(2 - \frac{2}{3} e^{-z} (3 + 3 z + z^2)\right) \ ,\nonumber \\
&& V_{T, PP}(z)  = - \frac{e^{-z} (1+z)}{24\pi z} \ , \nonumber \\
&& V_{T, MM}(z)  = \frac{ (1 + \kappa_1)^2\, m_\pi^2}{12 g_A^2 m_N^2} \frac{3}{4\pi z^3}\ .
\end{eqnarray}

The pion- and short-range potentials induced by dimension-nine $\Delta L=2$ operators are
\begin{eqnarray} \label{ourdef2}
V_{GT,\pi\pi}(z) &=&  -V_{GT,PP}
\qquad   V_{T,\, \pi \pi}(z) =  -V_{T,PP}
\nonumber \\
V_{GT, \pi N}(z) &=&  -\frac{1}{2}V_{GT,AP}
\qquad    V_{T,\pi N}(z) =  \frac{e^{-z} ( 3 + 3 z +z^2)}{12 \pi z^3}, \nonumber \\
V_{F,\, NN} &=& V_{GT,\, NN} =\delta^{(3)}( m_\pi \vec r).
\end{eqnarray}

\bibliography{bib2beta}
\end{document}